\documentclass[12pt]{article}
\RequirePackage{amsthm,amsmath,natbib}
\usepackage{amsmath, amsthm, amssymb}
\usepackage{verbatim,color,amssymb,epsfig}
\usepackage[usenames,dvipsnames]{xcolor}
\usepackage{multirow,booktabs}
 \usepackage{epsfig,natbib}
\usepackage{slashbox}
\usepackage{enumerate}
  \usepackage{float}
\usepackage{xr}
\def\boxit#1{\vbox{\hrule\hbox{\vrule\kern6pt
          \vbox{\kern6pt#1\kern6pt}\kern6pt\vrule}\hrule}}

\def\trans{^{\rm T}}

\numberwithin{equation}{section}\Large
\theoremstyle{plain}

\newtheorem{Theorem}{\underline{\bf Theorem}}[section]

\begin{document}

\thispagestyle{empty}
\baselineskip=28pt
\begin{center}
{\LARGE{\bf Functional regression with multivariate responses}}
\end{center}
\vskip 10mm

\baselineskip=14pt
\vskip 2mm
\begin{center}
 Ruiyan Luo\\
\vskip 2mm
Division of Epidemiology and Biostatistics, Georgia State University School of Public Health, 140 Decatur Street, Atlanta, GA 30303\\
rluo@gsu.edu\\
\hskip 5mm\\
Xin Qi\\
\vskip 2mm
Department of Mathematics and Statistics, Georgia State University, 25 Park Place, Atlanta, GA 30303-3083\\
xqi3@gsu.edu \\
\end{center}
 \pagenumbering{arabic}
\newlength{\gnat}
\setlength{\gnat}{22pt}
\baselineskip=\gnat

\begin{abstract}
We consider the functional regression model with multivariate response and functional predictors. Compared to fitting each individual response variable separately, taking advantage of the correlation between the response variables can improve the estimation and prediction accuracy. Using information in both functional predictors and multivariate response, we identify the optimal decomposition of the coefficient functions for prediction in population level. Then we propose methods to estimate this decomposition and fit the regression model for the situations of a small and a large number $p$ of functional predictors separately. For a large $p$,  we propose a simultaneous smooth-sparse penalty which can both make curve selection and improve estimation and prediction accuracy. We provide the asymptotic results when both the sample size and the number of functional predictors go to infinity. Our method can be applied to models with thousands of functional predictors and has been implemented in the \texttt{R} package \texttt{FRegSigCom}.

\end{abstract}
 \vspace{3mm}
 \noindent\underline{\bf Short Title}: Functional regression with multivariate response

\section{Introduction}\label{sec1}
 
Increasing interest has been shown in simultaneously predicting several quantities from a common set of predictor curves. For example, people want to predict the concentrations of several nutrition ingredients based on the near-infrared (NIR) spectrum of a corn sample. We consider the following linear regression model with multivariate response and functional predictors (MSOF),
 \begin{align}
\mathbf{Y}=\boldsymbol{\mu}+\sum_{j=1}^p\int_0^1 X_j(t)\mathbf{b}_j(t) dt+\boldsymbol{\varepsilon}=\boldsymbol{\mu}+\int \mathbb{B}(t)\trans\mathbf{X}(t)dt+\boldsymbol{\varepsilon},
\label{msof.eq}
\end{align} 
  where $\mathbf{Y}=(Y_1,\ldots,Y_m)\trans$ is an $m$-dimensional random vector and correlations usually exist  between the $Y_i$'s. The functional predictors $\mathbf{X}(t)=(X_1(t),\ldots,X_p(t))\trans$ are random functions and  without loss of generality, we assume that $X_j(t)$'s are defined in $[0,1]$ and have mean zero. Correlation can exist between $X_j(t)$'s.  The coefficient matrix $\mathbb{B}(t)=\left(\mathbf{b}_1(t), \ldots, \mathbf{b}_p(t)\right)\trans$ is $p\times m$, where $\mathbf{b}_j(t)=(b_{j1}(t), \ldots, b_{jm}(t))\trans$ is $m$-dimensional.  The integral of a vector or a matrix of functions is defined element-wise. The vectors $\boldsymbol{\mu}=(\mu_1,\ldots, \mu_m)\trans$ and $\boldsymbol{\varepsilon}=(\varepsilon_1,\ldots, \varepsilon_m)\trans$ represent  the intercept and random noise, respectively.  We assume that $\boldsymbol{\varepsilon}$ has mean zero and is independent of $\mathbf{X}(t)$.

Various methods have been proposed for scalar-on-function (SOF) linear regression model which is the special case of $\eqref{msof.eq}$ with $m=1$.   
We provide a brief review of these methods, and more thorough review can be found in \cite{morris2015functional}, \cite{wang2016functional}, and \cite{reiss2017methods}.  
A common general approach to fit the SOF model is to expand the coefficient functions (and probably the predictor curves) using a set of basis functions, and convert the SOF model to a linear model with multiple scalar predictors. Two popular ways exist to choose basis functions. First, the forms of the basis functions do not depend on the data, such as B-spline, P-spline, and Fourier basis. This approach has been adopted by \citet{ramsay1991some}, \citet{marx1999generalized}, \cite{cardot2003spline}, \cite{Ramsay-Silverman-2005}, \citet{crambes2009smoothing}, \cite{crainiceanu2010bayesian}, \cite{goldsmith2011penalized}, etc. 
Second, basis functions are adaptively chosen from data, such as the functional principal component regression (FPCR) \citep{Rice-Silverman-1991, Silverman-1996, cardot1999functional, yao2005functional, hall2006assessing, reiss2007functional, hall2007methodology,  hilgert2013minimax, li2013selecting, hosseini2013cross}, functional partial least squares (FPLS) \citep{preda2005pls, reiss2007functional, escabias2007functional,  kramer2008penalized,  aguilera2010using, delaigle2012methodology, hilgert2013minimax, li2013selecting, hosseini2013cross}, and functional sliced
inverse regression \citep{ferre2003functional, ferre2005smoothed}. FPCR, using the covariance function of the predictor curves to get the basis functions, are extensively studied and have several variants \citep{fengler2003dynamics, kneip2001inference, benko2009common, di2009multilevel, crainiceanu2009generalized, greven2011longitudinal}. 
 FPLS, using information in both response and predictors, obtains the basis functions by maximizing the covariance between the response and functional predictors.  
 When $p$ is large, hybrid methods are proposed which combine basis expansion and variable selection.   \cite{zhu2009functional}, \cite{lian2013shrinkage} and \cite{kong2016partially} perform FPCA on each functional predictor, and use PC scores as new scalar predictors to fit the model by imposing the group Lasso or group SCAD penalty.  \citet{kong2013classical}, \citet{swihart2014restricted} and \citet{collazos2016consistent} use hypothesis testing to select functional variables. 
Wavelet-based methods \citep{brown2001bayesian, zhao2012wavelet, zhao2015wavelet, reiss2015wavelet}  first transform the functional predictors to the wavelet space, and then use the wavelet coefficients as predictors to fit the model  with various sparse regularization or priors imposed.  
 In addition to the linear model, nonlinear SOF models have been considered, such as the functional single-index model \citep{eilers2009multivariate, ait2008cross, ferraty2011estimation}, the functional multiple-index model \citep{james2005functional, chen2011single, ferraty2013functional},  the functional nonparametric model \citep{ferraty2002functional, Ferraty-vieu-2006, rachdi2007nonparametric, shang2013bayesian, zhang2014sampling}, and so on. 

 For the MSOF regression, only a few methods have been proposed to jointly fit all coordinates of the multivariate response. \cite{matsui2008multivariate} expand both the predictor curves and coefficient functions using Gaussian basis functions, and transform the functional linear model to a linear model with multivariate response and multiple scalar predictors which is fitted using maximum likelihood estimation. \cite{wang2017gaussian} assume the model $\mathbf{Y}=\mathbf{f}(X)+\boldsymbol{\varepsilon}$, where $\mathbf{Y}$ is a multivariate response and $\mathbf{f}$ is an unknown multivariate function of $X(t)$. FPCA is first conducted on the multivariate response to de-correlate, and each PC score is separately used as a new response and fitted using the Gaussian process regression \citep{o1978curve, rasmussen2006gaussian}.   \cite{chaouch2013nonparametric} consider the nonparametric model with multivariate response and functional predictors, and propose an $L_1$-median estimation method.
 
For multivariate response, one can separately fit a SOF regression model for each coordinate of $\mathbf{Y}$ on $\mathbf{X}(t)$ using various methods. However, this approach ignores the correlation between the coordinates of $\mathbf{Y}$, which impairs the prediction accuracy and computational efficiency. 
One way to take account of the correlation in $\mathbf{Y}$ is to consider the decomposition of $\mathbb{B}(t)$ of the form $\sum_{k=1}^K\boldsymbol{\gamma}_k(t)\mathbf{v}_k\trans$, where $\boldsymbol{\gamma}_k(t)=(\gamma_{k1}(t), \ldots, \gamma_{kp}(t))\trans$ is a $p$-dimensional vector of functions, $\mathbf{v}_k$ is an $m$-dimensional vector, and $K$ is called the number of components. The estimates in both FPCR and FPLS belong to this type of decompositions. 
Aiming to estimate the regression function $\boldsymbol{\mu}_{\mathbf{Y}|\mathbf{X}}= \mathrm{E}[\mathbf{Y}|X_1(t),\ldots,X_p(t)]=\boldsymbol{\mu}+\sum_{j=1}^p\int X_j(t)\mathbf{b}_j(t)\trans dt$ and make prediction, we identify the optimal decomposition of this type in population level, and characterize this decomposition using a generalized eigenvalue problem which provides an efficient way to estimate this decomposition and fit the model in practice. The stronger correlation exists between the coordinates of $\mathbf{Y}$, the fewer components are needed to provide good estimation of the regression function and good prediction. So our methods take advantage of the correlation structure in the response to make efficient dimension reduction. We propose different regularization for models with a small number $p$ of functional predictors and a large $p$, separately. When $p$ is large, we propose a simultaneous sparse-smooth penalty which improves the estimation and prediction accuracy by selecting a small number of predictor curves in the fitted model. We provide asymptotic results as both $p$ and $n$ go to infinity.  Intensive simulation studies and two real data analysis show good performance of our methods, which we have implemented in the \texttt{R} package \texttt{FRegSigCom}.

 We start with the model with $p=1$ to introduce the optimal decomposition and our estimation method in Section~\ref{sec2}. We extend the method to models with $p>1$ in Section~\ref{sec3} and  consider computational issues in Section \ref{section_4}. We conduct simulation studies and real data analysis in Sections \ref{section_5} and \ref{section_6}, respectively.  
 We provide the proofs, algorithms and additional formulas and figures in supplementary materials.

\section{Regression with one functional predictor}\label{sec2}

 When $p=1$, the coefficient matrix $\mathbb{B}(t)$ in model  $\eqref{msof.eq}$  becomes an $m$-dimensional vector and is denoted as $\mathbf{b}(t)$. Then model $\eqref{msof.eq}$ becomes
\begin{align}
\mathbf{Y}=\boldsymbol{\mu}+\int_0^1 X(t)\mathbf{b}(t) dt+\boldsymbol{\varepsilon}.
\label{001}
\end{align}

\subsection{Optimal decomposition for $\mathbf{b}(t)$}\label{sec2.1}    

For model $\eqref{001}$, we consider all decompositions of $\mathbf{b}(t)$ of the form $\sum_{k=1}^K\gamma_k(t)\mathbf{v}_k$, where $\gamma_k(t)$'s are functions and $\mathbf{v}_k$'s are $m$-dimensional vectors. 
We will identify the optimal decomposition of this form for estimating the regression function $\boldsymbol{\mu}_{\mathbf{Y}|\mathbf{X}}=\boldsymbol{\mu}+\int_0^1 X(t)\mathbf{b}(t) dt$ and predicting the response in population level.  Noting that the estimates in FPCR and FPLS have this form of decomposition,  afterward we will compare the optimal one with those in FPCA and FPLS.

 Let  $\Sigma(s,t)={\mathrm{E}}[X(s)X(t)]$ be the covariance function of $X(t)$ and define  
\begin{align}
\Gamma(s,t)&=\int_0^1\int_0^1\Sigma(s,s^\prime)\mathbf{b}(s^\prime)\trans\mathbf{b}(t^\prime)\Sigma(t^\prime,t)ds^\prime dt^\prime.\label{33003}
\end{align} 
  Let $\alpha_1(t)$ be the solution to 
\begin{align}
&\max_{\alpha(t) }  \int_0^1\int_0^1\alpha(s)\Gamma(s,t)\alpha(t)dsdt, \quad \text{\rm subject to}  \int_0^1\int_0^1\alpha(s)\Sigma(s,t)\alpha(t)dsdt=1, \label{2220.1}
\end{align}	
and ${\alpha}_k(t)$, $1<k\le K$, be the solution to
\begin{align}
&\max_{\alpha(t) }  \int_0^1\int_0^1\alpha(s)\Gamma(s,t)\alpha(t)dsdt, \quad \text{\rm subject to}  \int_0^1\int_0^1\alpha(s)\Sigma(s,t)\alpha(t)dsdt=1\notag\\ 
& \qquad\qquad\qquad \text{and } \int_0^1\int_0^1\alpha_{k'}(s)\Sigma(s,t)\alpha(t)dsdt=0, \quad 1\le k' \le k-1.\label{2220.k}
\end{align}	
The $\eqref{2220.1}$ and $\eqref{2220.k}$ define a sequence of generalized eigenvalue problems with solutions $\{{\alpha}_k(t), 1\le k \le K\}$ called the eigenfunctions. As both $\Sigma(s,t)$ and $\Gamma(s,t)$ are symmetric nonnegative definite kernel functions, the maximum values of  $\eqref{2220.1}$ and $\eqref{2220.k}$ are nonnegative and denoted as $\sigma_1^2 \ge \sigma_2^2 \ge \cdots \ge \sigma_K^2$. We call $\sigma_k^2$ the eigenvalue corresponding to the $k$-th eigenfunction $\alpha_k(t)$. 
Define an $m$-dimensional vector $\mathbf{w}_k=\int_0^1\int_0^1\alpha_k(s)\Sigma(s,t)\mathbf{b}(t) dsdt$ for $k\ge 1$.  In the following theorem, we show that $\sum_{k=1}^K\alpha_k(t)\mathbf{w}_k$ is the optimal decomposition. 

\begin{Theorem}\label{theorem_1}
 Let $K_0$ denote the total number of positive eigenvalues of the generalized eigenvalue problem $\eqref{2220.k}$ and $\sigma_1^2\ge \cdots \ge \sigma_{K_0}^2>0= \sigma_{K_0+1}^2 \cdots$ denote the decreasingly sorted eigenvalues. The $\alpha_k(t)$'s and $\mathbf{w}_k$'s are defined as above. \\
(1). $K_0$ is less than or equal to $m$, and $(\sigma_k^2, \mathbf{w}_k)$ is also the $k$-th  eigenvalue-eigenvector pair of the covariance matrix of $\boldsymbol{\mu}_{\mathbf{Y}|\mathbf{X}}$, for $1\le k \le K_0$.\\
(2).  The vector $\boldsymbol{\mu}+\int_0^1 X(t)\{\sum_{k=1}^K\alpha_k(t)\mathbf{w}_k\}dt$ has the smallest mean squared error in approximating the regression function $\boldsymbol{\mu}_{\mathbf{Y}|\mathbf{X}}$. Specifically, we have
\begin{align}
 &  {\mathrm{E}}\left[\left\|\boldsymbol{\mu}_{\mathbf{Y}|\mathbf{X}}-\boldsymbol{\mu}- \int_0^1 X(t)\left\{\sum_{k=1}^K\alpha_k(t)\mathbf{w}_k\right\}dt\right\|_2^2\right]=\sum_{k=K+1}^{K_0}\sigma_k^2 \label{300011}\\
=&\min_{\substack{\boldsymbol{\nu}, \gamma_k(t), \mathbf{v}_k,\\ 1\le k\le K}}\left\{{\mathrm{E}}\left[\left\|\boldsymbol{\mu}_{\mathbf{Y}|\mathbf{X}}-\boldsymbol{\nu}- \int_0^1 X(t)\left\{\sum_{k=1}^K{\gamma}_k(t)\mathbf{v}_k\right\}dt\right\|_2^2\right]\right\},\notag
\end{align}
where the minimum is taken over all square integrable functions  $\gamma_k(t)$'s and all $m$-dimensional  vectors $\boldsymbol{\nu}$ and $\mathbf{v}_k$'s. \\
(3).  For a new observation $(X_{\rm new}(t), \mathbf{Y}_{\rm new})$, the predicted response vector given by $\boldsymbol{\mu}+\int_0^1 X_{\rm new}(t)\{\sum_{k=1}^K\alpha_k(t)\mathbf{w}_k\}dt$ has the smallest prediction error as follows,  
\allowdisplaybreaks\begin{align}
 & {\mathrm{E}}\left[\left\|\mathbf{Y}_{\rm new}-\boldsymbol{\mu}- \int_0^1 X_{\rm new}(t)\left\{\sum_{k=1}^K\alpha_k(t)\mathbf{w}_k\right\}dt\right\|_2^2\right]=\sum_{k=K+1}^{K_0}\sigma_k^2+{\mathrm{E}}[\|\boldsymbol{\varepsilon}_{\rm new}\|_2^2]\label{300001}\\
=&\min_{\substack{\boldsymbol{\nu}, \gamma_k(t), \mathbf{v}_k,\\ 1\le k\le K}}\left\{{\mathrm{E}}\left[\left\|\mathbf{Y}_{\rm new}-\boldsymbol{\nu}- \int_0^1 X_{\rm new}(t)\left\{\sum_{k=1}^K{\gamma}_k(t)\mathbf{v}_k\right\}dt\right\|_2^2\right]\right\},\notag
\end{align}  
where $\boldsymbol{\varepsilon}_{\rm new}$ is the noise vector in $\mathbf{Y}_{\rm new}$ and is independent of $X_{\rm new}(t)$ and the original variables $(X(t), \mathbf{Y})$.
\end{Theorem}

\begin{figure}[h]
\includegraphics[height=2.5in,width=6.5in]{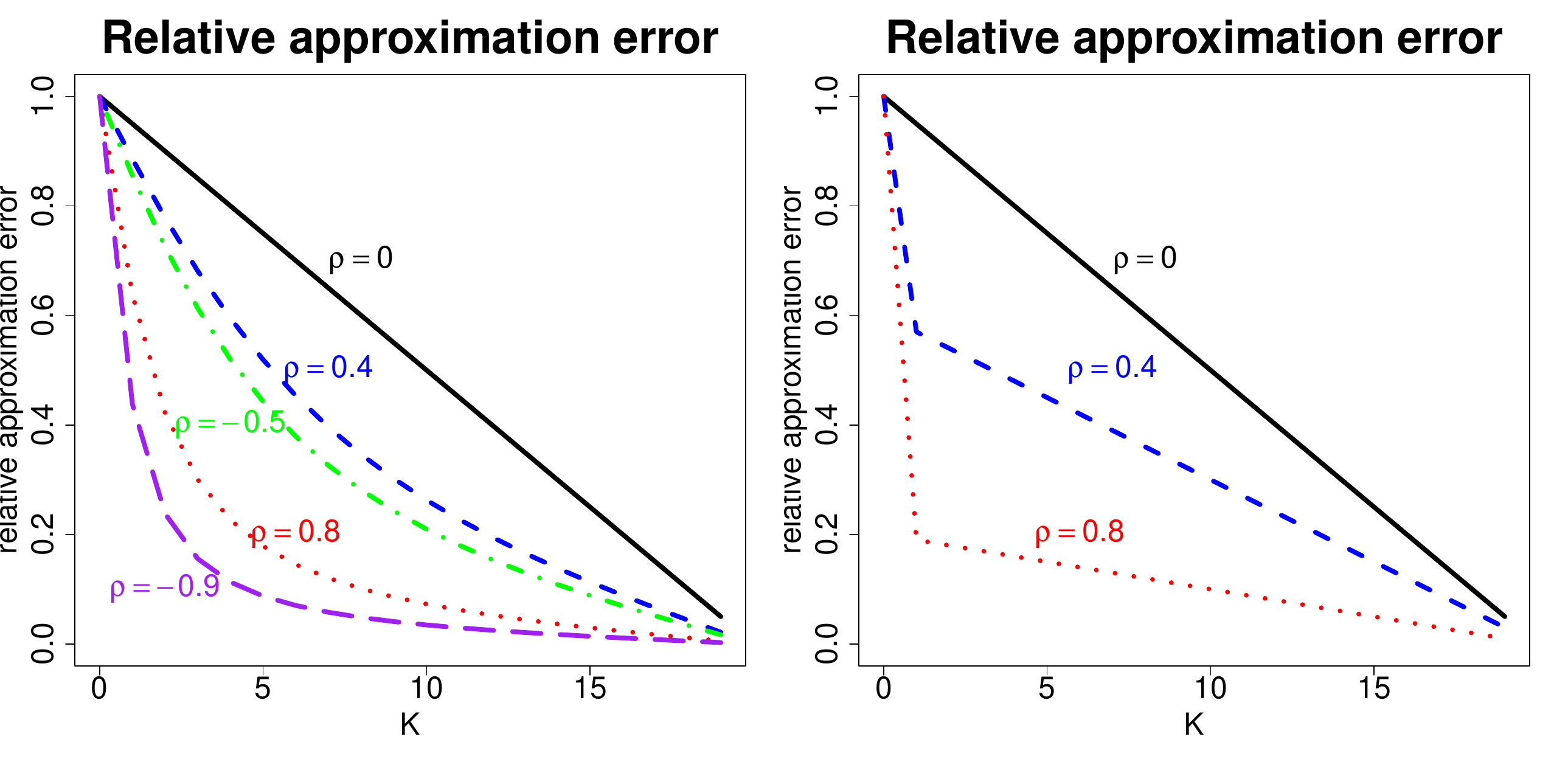}
    \caption{\label{fig_1} \baselineskip=10pt Two examples for the relative approximation errors of the optimal decomposition of $\mathbf{b}(t)$ versus the number $K$ of components, where different correlation levels indicated by $\rho$ are considered. The left plot is for the covariance matrix of $\boldsymbol{\mu}_{\mathbf{Y}|\mathbf{X}}$ with the $(i,j)$-th entry equal to $\rho^{|i-j|}$; the right plot is for the  covariance matrix with diagonal elements equal to 1 and off-diagonal elements equal to $\rho$.
}
\end{figure}

Theorem \ref{theorem_1} shows that given $K$, the goodness of $\boldsymbol{\mu}+ \int_0^1 X(t) \left\{\sum_{k=1}^K  {\alpha}_{k}(t)\mathbf{w}_k\right\}  dt$ in approximating $\boldsymbol{\mu}_{\mathbf{Y}|\mathbf{X}}$ can be measured by $\sum_{k>K} \sigma_k^2$. Therefore, the faster is the decay of $\sigma_1^2\ge\sigma_2^2\ge\cdots $,  a smaller number $K$ of components is needed to provide an adequate approximation to $\boldsymbol{\mu}_{\mathbf{Y}|\mathbf{X}}$. The decay rate of $\sigma_k^2$'s is affected by the correlation between the coordinates in $\boldsymbol{\mu}_{\mathbf{Y}|\mathbf{X}}$. Generally speaking, stronger correlation between the coordinates in $\boldsymbol{\mu}_{\mathbf{Y}|\mathbf{X}}$ leads to a faster decrease of $\sigma_k^2$. As an example, we take $m=20$ and consider two cases for the covariance matrix of $\boldsymbol{\mu}_{\mathbf{Y}|\mathbf{X}}$. \underline{\it Case 1}: the $(i,j)$-th entry of the covariance matrix  is $\rho^{|i-j|}$; \underline{\it Case 2}: all the diagonal elements of the covariance matrix equal to 1 and all others are $\rho$, where in both cases, $-1 \le \rho \le 1$. A larger  value of $|\rho|$ indicates stronger correlation between the coordinates of $\boldsymbol{\mu}_{\mathbf{Y}|\mathbf{X}}$. We plot the {\it relative approximation errors}, $\sum_{k=K+1}^m\sigma_k^2/\sum_{k=1}^m\sigma_k^2$, versus $K$ for the two cases, respectively, in Figure \ref{fig_1} which shows that a larger value of $|\rho|$ leads to a faster decay of the relative approximation error in both cases.

Assuming that the expectation of the predictor curve is zero, the squared prediction error in $\eqref{300001}$ also gives the sum of variances of all the coordinates of prediction  vector. The first equality in $\eqref{300001}$ shows that the squared prediction error consists of two parts: $\sum_{k=K+1}^{K_0}\sigma_k^2$ and ${\mathrm{E}}[\|\boldsymbol{\varepsilon}_{\rm new}\|_2^2]$. The second part is the variance of the noise vector $\boldsymbol{\varepsilon}_{\rm new}$ which is independent of $X_{\rm new}(t)$ and the original data $(X(t), \mathbf{Y})$. It can not be reduced when prediction is based on $X_{\rm new}(t)$ and $(X(t), \mathbf{Y})$. The first part $\sum_{k=K+1}^{K_0}\sigma_k^2$ is due to the truncation of the decomposition $\sum_{k=1}^{K_0}\alpha_k(t)\mathbf{w}_k$ after the first $K$ terms. It can be reduced by increasing the number $K$ of components. However, in practice, we have to estimate $\alpha_k(t)$'s and $\mathbf{w}_k$'s. So in addition to the truncation error and the random error, the prediction error in practice also includes the estimation error of $\alpha_k(t)$'s and $\mathbf{w}_k$'s.  Despite the reduction of the truncation error, a larger $K$ leads to more functions to be estimated, which brings additional estimation error. In practice, we have to choose an appropriate $K$ to achieve a balance between the prediction error due to truncation and that due to estimation. 
 
\begin{figure}[h]
\includegraphics[height=3.5in,width=6.5in]{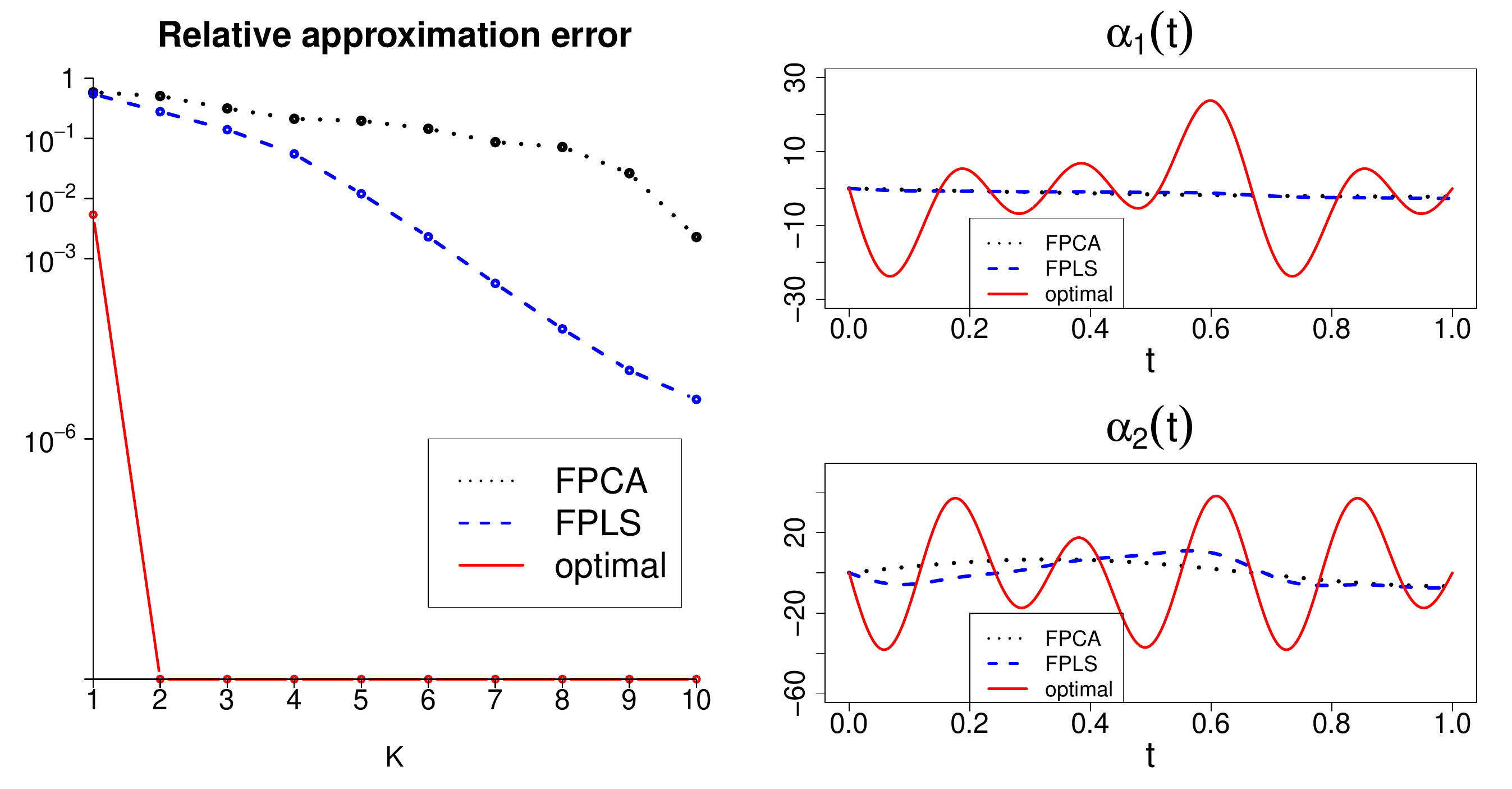}
    \caption{\label{fig_15} \baselineskip=10pt  Comparison of the components  $\alpha_k(t)$ obtained by 3 methods:  FPCA, FPLS and our optimal decomposition. For comparison, all components are scaled so that $\int_0^1\int_0^1\alpha(s)\Sigma(s,t)\alpha(t)dsdt=1$.
		{\it Top-left}: the first component $\alpha_1(t)$. {\it Top-right}: the second component. {\it Bottom-left}: the third component. {\it Bottom-right}: the relative approximation error $\boldsymbol{\mu}_{\mathbf{Y}|\mathbf{X}}$ versus the number $K$ of components for the three methods.}
\end{figure}

Two popular dimension reduction methods, FPCR and FPLS, also generate the decompositions of the form $\sum_{k=1}^K\gamma_k(t)\mathbf{v}_k$ for $\mathbf{b}(t)$. In FPCR, $\gamma_k(t)$'s are eigenfunctions of the covariance of the predictor curve.
 The FPLS extends the idea of PLS to build orthogonal components that maximize their covariance with response. There are different variants of PLS \citep{wold1973nonlinear, wold1975path,  ter1998objective, boulesteix2006partial}, and hence different versions of FPLS have been proposed \citep{goutis1996partial, reiss2007functional, kramer2008penalized, kramer2011degrees, delaigle2012methodology}. For example, the FPLS implemented in software \texttt{R} is based on the work of \cite{kramer2008penalized} and \cite{kramer2011degrees} that finds uncorrelated latent variables to maximize the squared covariance between $\mathbf{Y}$ and the latent variables where the length of loading vectors equal to 1. In the population level, this FPLS solves
 \begin{align}
\max_{\alpha(t) } \quad & \int_0^1\int_0^1\alpha(s)\Gamma(s,t)\alpha(t)dsdt, \label{pls.eq}\\ 
 \text{\rm subject to}\quad   &\|\alpha\|_{L^2}=1,\quad \text{and } \int_0^1\int_0^1\alpha_{k'}(s)\Sigma(s,t)\alpha(t)dsdt=0, \quad 1\le k' \le k-1.\notag
\end{align}	
Problems $\eqref{2220.k}$ and $\eqref{pls.eq}$ share a common objective function which can be expressed as 
 \small\begin{align}
 &\int_0^1\int_0^1\alpha(s)\Gamma(s,t)\alpha(t)dsdt =\mathrm{E}\left[\int_0^1 X(t)\alpha_k(t) dt (\mathbf{Y}-\boldsymbol{\mu})\right]\trans \mathrm{E}\left[\int_0^1 X(t)\alpha_k(t) dt (\mathbf{Y}-\boldsymbol{\mu})\right], \label{56901}
\end{align}
\normalsize
 the sum of squared covariance between $\int_0^1 X(t)\alpha_k(t) dt$ and each response coordinate. 
The common restriction $0=\int_0^1\int_0^1\alpha_{k'}(s)\Sigma(s,t)\alpha(t)dsdt=\mathrm{Cov}(\int_0^1 X(t)\alpha_k(t) dt, \int_0^1 X(t)\alpha_{k'}(t) dt)$ implies that $\int_0^1 X(t)\alpha_k(t) dt$ and  $\int_0^1 X(t)\alpha_{k'}(t) dt$ are uncorrelated for $k \ne k'$. But FPLS in $\eqref{pls.eq}$ requires that  $\alpha_k$ has unit length, and the region $\{\alpha(t): \|\alpha\|_{L^2}=1\}$ for $\eqref{pls.eq}$ is the unit sphere in $L^2[0,1]$; while $\eqref{2220.k}$ requires that $\int_0^1 X(t)\alpha_k(t) dt$ has unit variance as 
\[
\mathrm{Var}\left(\int_0^1 X(t)\alpha_k(t) dt\right)=\mathrm{E}\left[\left\{\int_0^1 X(t)\alpha_k(t) dt\right\}^2\right]= \int_0^1\int_0^1\alpha(s)\Sigma(s,t)\alpha(t)dsdt=1,
\]
and the region $\{\alpha(t): \int_0^1\int_0^1\alpha(s)\Sigma(s,t)\alpha(t)dsdt=1\}$  is an ellipsoid. 
Hence the solutions to $\eqref{2220.k}$ and $\eqref{pls.eq}$ are usually different.  In Figure~\ref{fig_15}, we provide an example to illustrate the difference between the optimal decomposition and those for FPCR and FPLS. Let $X(t)$ be the Brownian motion in $[0,1]$, and its covariance function is $\Sigma(t^\prime,t)=\min(t^\prime,t)$. Let $\mathbf{b}(s)=(b_1(s), \ldots, b_5(s))\trans$ be a five-dimensional vector with $b_k(s)=\sin(k\pi t)+2\sin((k+1)\pi t)$ for $1\le k\le 5$. We draw the first three component functions used in decomposition, and  the  relative approximation errors to $\boldsymbol{\mu}_{\mathbf{Y}|\mathbf{X}}$ versus the number $K$ of components for each of three decomposition methods in Figure \ref{fig_15} which shows that our components are quite different from those in FPCR and FPLS. For each $K$, the optimal decomposition leads to the smallest approximation error to $\boldsymbol{\mu}_{\mathbf{Y}|\mathbf{X}}$. When $K=1$, the approximation error of the optimal decomposition is less than half of any of the other two. 
 When $K=5$ (the dimension of the response), the approximation error of the optimal decomposition is zero, whereas the other two still have positive errors. Indeed, Theorem \ref{theorem_1} shows that the optimal decomposition can provide a perfect approximation to $\boldsymbol{\mu}_{\mathbf{Y}|\mathbf{X}}$ with $K$ not greater than the dimension of the response. This fact provides an upper bound for our choice of optimal $K$ in practice.

\subsection{Estimation of the optimal decomposition}\label{sec2.2}
   
Let $\{\mathbf{Y}_\ell=(Y_{\ell 1},\ldots,Y_{\ell m})\trans$, $X_\ell(t): 1\le \ell\le n\}$ be $n$ independent samples from the model $\eqref{001}$, and $\boldsymbol{\varepsilon}_{\ell}$ denote the noise in $\mathbf{Y}_\ell$. Let $\overline{\mathbf{Y}}$ and  $\overline{X}(t)$ denote sample means, and $\widehat{\Sigma}(s,t)=\sum_{\ell=1}^n\left\{X_\ell(s)-\overline{X}(s)\right\}\left\{X_{\ell}(t)-\overline{X}(t)\right\}/n$ denote the sample covariance function of predictor. 
Let $\widehat{\Gamma}(s,t)=\left[\sum_{l=1}^n\left\{X_{l}(s)-\overline{X}(s)\right\}\{\mathbf{Y}_{l}-\overline{\mathbf{Y}}\}\trans\right]\left[\sum_{\ell^\prime=1}^n \{\mathbf{Y}_{\ell^\prime}-\overline{\mathbf{Y}}\}\left\{X_{\ell^\prime}(t)-\overline{X}(t)\right\}\right]/n^2$. To see that $\widehat{\Gamma}(s,t)$ is an estimator of $\Gamma(s,t)$, plugging $\mathbf{Y}_l=\boldsymbol{\mu}+\int_0^1 X_l(t)\mathbf{b}(t) dt + \varepsilon_{\ell}$ into the definition of $\widehat{\Gamma}(s,t)$,  we have
 \begin{align*}
  \widehat{\Gamma}(s,t)=& \int_0^1\int_0^1\widehat{\Sigma}(s,s^\prime)\mathbf{b}(s^\prime)\trans\mathbf{b}(t^\prime)\widehat{\Sigma}(t^\prime,t)ds^\prime dt^\prime \notag\\
&+ \frac{2}{n} \sum_{\ell^\prime=1}^n \{\boldsymbol{\varepsilon}_{\ell^\prime}-\overline{\boldsymbol{\varepsilon}}\}\left\{X_{\ell^\prime}(t)-\overline{X}(t)\right\}\int_0^1\widehat{\Sigma}(s,s^\prime)\mathbf{b}(s^\prime)\trans ds^\prime  \notag\\
& + \frac{1}{n^2}\left[\sum_{l=1}^n\left\{X_{l}(s)-\overline{X}(s)\right\}\{\boldsymbol{\varepsilon}_{l}-\overline{\boldsymbol{\varepsilon}}\}\trans\right]\left[\sum_{\ell^\prime=1}^n \{\boldsymbol{\varepsilon}_{\ell^\prime}-\overline{\boldsymbol{\varepsilon}}\}\left\{X_{\ell^\prime}(t)-\overline{X}(t)\right\}\right],
\end{align*}
where the last two terms converge in probability to zero as $n\to\infty$ due to the law of large numbers and the independence of $X_l(s)$'s and $\boldsymbol{\varepsilon}_{l}$'s.  So $\widehat{\Gamma}(s,t)$ is a consistent estimator of $\Gamma(s,t)$.  
As the generalized eigenvalue problem $\eqref{2220.k}$ is equivalent to the following Rayleigh quotient maximization problem,
\begin{align*}
&\max_{{\alpha}(t) }  \frac{\int_0^1\int_0^1\alpha(s)\Gamma(s,t){\alpha}(t)dsdt}{\int_0^1\int_0^1\alpha(s)\Sigma(s,t){\alpha}(t)dsdt}, \quad \text{\rm subject to}  \int_0^1\int_0^1\alpha(s)\Sigma(s,t){\alpha}(t)dsdt=1\notag\\ 
& \qquad\qquad\qquad \text{and } \int_0^1\int_0^1\alpha_{k'}(s)\Sigma(s,t){\alpha}(t)dsdt=0, \quad 1\le k'\le k-1,
\end{align*}
we propose to get the estimate $\widehat{\alpha}_k(t)$ of $\alpha_k(t)$ in the optimal decomposition by solving
\begin{align}
&\max_{{\alpha}(t) }  \frac{\int_0^1\int_0^1\alpha(s)\widehat{\Gamma}(s,t){\alpha}(t)dsdt}{\int_0^1\int_0^1\alpha(s)\widehat{\Sigma}(s,t){\alpha}(t)dsdt+P({\alpha})}, \quad \text{\rm subject to}  \int_0^1\int_0^1\alpha(s)\widehat{\Sigma}(s,t){\alpha}(t)dsdt=1\notag\\ 
& \qquad\qquad\qquad \text{and } \int_0^1\int_0^1\widehat{\alpha}_{k'}(s)\widehat{\Sigma}(s,t){\alpha}(t)dsdt=0, \quad 1\le k'\le k-1,\label{112220}
\end{align}
 where  $P(\alpha)=\tau\left(\|\alpha\|_{L^2}^2+\eta\|\alpha^{\prime\prime}\|^2_{L^2}\right)$,  $\alpha^{\prime\prime}(t)$ represents the second derivative of $\alpha(t)$, and  $\|\cdot\|_{L^2}$ denotes the $L^2$ norm of a function. The way we impose the penalty $P(\alpha)$ in $\eqref{112220}$ has been used in functional data analysis. For example, in the regularized FPCA (Section 9.3.1 of \citet{Ramsay-Silverman-2005}), a penalized objective function, $\int\int\xi(s)\widehat{\Sigma}(s,t)\xi(t)dsdt/(\|\xi\|^2_{L^2}+\lambda\|\xi''\|^2_{L^2})$, is maximized over all smooth function $\xi(s)$. Imposing penalty in the denominator of the objective function, we will have a smaller value of the objective function for a rougher $\xi(s)$ with a large value of $\|\xi''\|^2_{L^2}$. So the maximizer of this objective function is a smooth function, and the larger is $\lambda$, the smoother is the maximizer.  In our penalty $P(\alpha)$, the term $\|\alpha^{\prime\prime}\|^2_{L^2}$ is the smoothness penalty and plays the same role as that in the regularized FPCA. On the other hand, as $\widehat{\Sigma}(s,t)$ is estimated from a finite number of samples,  it has only a finite number of positive eigenvalues.  If we only impose the smoothness  penalty in $\eqref{112220}$, the solution to $\eqref{112220}$ may not be unique and the solution $\widehat{\alpha}_k(s)$ can have a much larger norm than $\alpha_k(t)$, just like the overfitting problems due to multicollinearity in classic linear model. Motivated by the ridge penalty which is used to shrink the estimated coefficients and alleviate the effect of multicollinearity, we add the term $\tau\|\alpha\|_{L^2}^2$ in our penalty to control the norm of $\alpha$.  The parameter $\eta$  tunes the relative importance between $\|\alpha\|_{L^2}^2$ and $\|\alpha^{\prime\prime}\|^2_{L^2}$ in the penalty.

Suppose that we have obtained the estimates $\widehat{\alpha}_k(t)$, $1\le k\le K$. To estimate the intercept vector $\boldsymbol{\mu}$ and the vectors $\mathbf{w}_k$'s in the optimal decomposition, we define  $\widehat{{T}}_{\ell,k}=\int_0^1\{X_\ell(t)-\overline{X}(t)\}\widehat{\alpha}_k(t)dt$ for  $1 \le \ell \le n$ and $1 \le k \le K$, and let $\widehat{\mathbf{T}}_k=(\widehat{{T}}_{1,k}, \ldots,\widehat{{T}}_{n,k})\trans$. Then we  get the estimates $\{\widehat{\boldsymbol{\mu}}, \widehat{\mathbf{w}}_1, \ldots, \widehat{\mathbf{w}}_K\}$ by regressing $\mathbf{Y}_{\ell}$'s on the new scalar predictors $\widehat{\mathbf{T}}_k$'s using the following least squares problem,
\begin{align}
  \min_{\substack{\boldsymbol{\nu},  \mathbf{v}_k, 1\le k\le K}}\left\{\sum_{\ell=1}^n\left\|\mathbf{Y}_{\ell}-\boldsymbol{\nu}-  \sum_{k=1}^K\widehat{T}_{\ell,k}\mathbf{v}_k \right\|_2^2\right\},\label{3671}
\end{align}
where the minimum is taken over all $m$-dimensional vectors $\boldsymbol{\nu}$ and $\mathbf{v}_k$'s. As the  $\widehat{\mathbf{T}}_k$'s are orthogonal, the solution to $\eqref{3671}$ can be explicitly given by $\widehat{\boldsymbol{\mu}}= \overline{\mathbf{Y}}$ and $\widehat{\mathbf{w}}_k=\sum_{\ell=1}^n\widehat{T}_{\ell,k}\mathbf{Y}_{\ell}/n$ for $1\le k\le K$. Finally, we  estimate $\boldsymbol{\mu}_{\mathbf{Y}|\mathbf{X}}$  by  $\widehat{\boldsymbol{\mu}}_{\mathbf{Y}|\mathbf{X}}=\widehat{\boldsymbol{\mu}}+\int_0^1X(t)\{\sum_{k=1}^K\widehat{\alpha}_k(t)\widehat{\mathbf{w}}_k\}dt$. Given a new observation $X_{\rm new}(t)$ of the functional predictor, we predict the response vector as $\mathbf{Y}_{\rm pred}=\widehat{\boldsymbol{\mu}}+\int_0^1X_{\rm new}(t)\{\sum_{k=1}^K\widehat{\alpha}_k(t)\widehat{\mathbf{w}}_k\}dt$. 
 
\section{Regression with multiple functional predictors}\label{sec3}
For model $\eqref{msof.eq}$ with multiple functional predictors ($p>1$), we first provide the optimal decomposition for the $p\times m$ matrix $\mathbb{B}(t)$ of coefficient functions in the population level. Then we estimate this decomposition separately for two cases: a small $p$ and a large $p$. 

	\subsection{Optimal decomposition for $\mathbb{B}(t)$}\label{sec3.1}
Extending the decomposition in Section~\ref{sec2.1}, for model $\eqref{msof.eq}$ with $p>1$, we consider the decomposition of $\mathbb{B}(t)$ of the form $\sum_{k=1}^K\boldsymbol{\gamma}_k(t)\mathbf{v}_k\trans$, where $\boldsymbol{\gamma}_k(t)=(\gamma_{k1}(t), \ldots, \gamma_{kp}(t))\trans$ is a $p$-dimensional vector of functions, and $\mathbf{v}_k$ is an $m$-dimensional vector. To find the optimal one of such form, we define two $p\times p$ matrices of kernel functions
\begin{align}
\boldsymbol{\Sigma}(s,t)=\mathrm{E}[\mathbf{X}(s)\mathbf{X}(t)\trans], \quad \boldsymbol{\Gamma}(s,t)=\int_0^1\int_0^1\boldsymbol{\Sigma}(s,s^\prime)\mathbb{B}(s^\prime)\mathbb{B}(t^\prime)\trans\boldsymbol{\Sigma}(t^\prime,t)ds^\prime dt^\prime, \label{3673}
\end{align} 
  which respectively generalize the scalar functions $\Sigma(s,t)$ and $\Gamma(s,t)$ in Section \ref{sec2.1}. Here  $\boldsymbol{\Sigma}(s,t)$ is the matrix of covariance functions of $\mathbf{X}(t)$. Let $\boldsymbol{\alpha}_k(t)=(\alpha_{k1}(t), \ldots, \alpha_{kp}(t))\trans$, $k \ge 1$, denote eigenfunctions of the following generalized eigenvalue problem: 
\begin{align}
&\max_{\boldsymbol{\alpha}(t) }  \int_0^1\int_0^1\boldsymbol{\alpha}(s)\trans\boldsymbol{\Gamma}(s,t)\boldsymbol{\alpha}(t)dsdt, \quad \text{\rm subject to}  \int_0^1\int_0^1\boldsymbol{\alpha}(s)\trans\boldsymbol{\Sigma}(s,t)\boldsymbol{\alpha}(t)dsdt=1\notag\\ 
& \qquad\qquad\qquad \text{and } \int_0^1\int_0^1\boldsymbol{\alpha}_{k'}(s)\trans\boldsymbol{\Sigma}(s,t)\boldsymbol{\alpha}(t)dsdt=0, \quad 1\le k'\le k-1,
\label{opt.large.p}
\end{align}
and define  the $m$-dimensional  vectors $\mathbf{w}_k=\int_0^1\int_0^1\mathbb{B}(t)\trans\boldsymbol{\Sigma}(s,t)\boldsymbol{\alpha}_k(s)dsdt$ for $k\ge 1$. As in Theorem~\ref{theorem_1}, we can show that $\sum_{k=1}^K\boldsymbol{\alpha}_{k}(t)\mathbf{w}_k\trans$ is the optimal decomposition of $\mathbb{B}(t)$ in approximating the regression function and prediction.
 
\subsection{Estimation of the optimal decomposition}\label{sec3.2}
Let $\mathbf{Y}_{\ell}=(Y_{\ell 1},\ldots,Y_{\ell m})\trans$ and $\mathbf{X}_\ell(t)=(X_{\ell 1}(t), \ldots, X_{\ell p}(t))\trans$, $1\le \ell\le n$, be $n$ independent samples from the model $\eqref{msof.eq}$. We estimate the matrices $\boldsymbol{\Sigma}(s,t)$ and $\boldsymbol{\Gamma}(s,t)$  by  
\begin{align}
 &\widehat{\boldsymbol{\Sigma}}(s,t)=\frac{1}{n} \sum_{\ell=1}^n\left\{\mathbf{X}_\ell(s)-\overline{\mathbf{X}}(s)\right\}\left\{\mathbf{X}_\ell(t)-\overline{\mathbf{X}}(t)\right\}\trans,\label{7}\\
&\widehat{\boldsymbol{\Gamma}}(s,t)=\frac{1}{n^2}\left[\sum_{l=1}^n\left\{\mathbf{X}_l(s)-\overline{\mathbf{X}}(s)\right\}\{\mathbf{Y}_{l}-\overline{\mathbf{Y}}\}\trans\right]\left[\sum_{\ell=1}^n \{\mathbf{Y}_\ell-\overline{\mathbf{Y}}\}\left\{\mathbf{X}_\ell(t)-\overline{\mathbf{X}}(t)\right\}\trans\right] \notag
\end{align} 
respectively, where $\overline{\mathbf{X}}(t)=\sum_{\ell=1}^n \mathbf{X}_{\ell}(t)/n$ and $\overline{\mathbf{Y}}=\sum_{\ell=1}^n \mathbf{Y}_{\ell}/n$. Our estimate $\widehat{\boldsymbol{\alpha}}_k(t)$ is the solution to the following penalized generalized eigenvalue problem,
\begin{align}
&\max_{\boldsymbol{\alpha}(t) }  \frac{\int_0^1\int_0^1\boldsymbol{\alpha}(s)\trans\widehat{\boldsymbol{\Gamma}}(s,t)\boldsymbol{\alpha}(t)dsdt}{\int_0^1\int_0^1\boldsymbol{\alpha}(s)\trans\widehat{\boldsymbol{\Sigma}}(s,t)\boldsymbol{\alpha}(t)dsdt+P(\boldsymbol{\alpha})}, \quad \text{\rm subject to}  \int_0^1\int_0^1\boldsymbol{\alpha}(s)\trans\widehat{\boldsymbol{\Sigma}}(s,t)\boldsymbol{\alpha}(t)dsdt=1\notag\\ 
& \qquad\qquad\qquad \text{and } \int_0^1\int_0^1\widehat{\boldsymbol{\alpha}}_{k'}(s)\trans\widehat{\boldsymbol{\Sigma}}(s,t)\boldsymbol{\alpha}(t)dsdt=0, \quad 1\le k'\le k-1,\label{12220}
\end{align}
 where $P(\boldsymbol{\alpha})$ is the penalty imposed on $\boldsymbol{\alpha}(s)=(\alpha_1(s),\ldots, \alpha_p(s))\trans$. We provide the form of $P(\boldsymbol{\alpha})$ for the following two situations separately.  

\noindent \underline{\it Situation 1: $p$ is relatively small.}  We consider the following penalty which is a direct extension of that in $\eqref{112220}$:
\begin{align}
& P(\boldsymbol{\alpha})=\tau \sum_{j=1}^p\|\alpha_j\|^2_\eta,\quad \text{where}\quad \|\alpha_j\|_\eta^2= \|\alpha_j\|_{L^2}^2+\eta\|\alpha^{\prime\prime}_j\|^2_{L^2}.\label{3677}
\end{align}
 
\begin{figure}[h]
\begin{center}
\includegraphics[height=2.5in,width=5.5in]{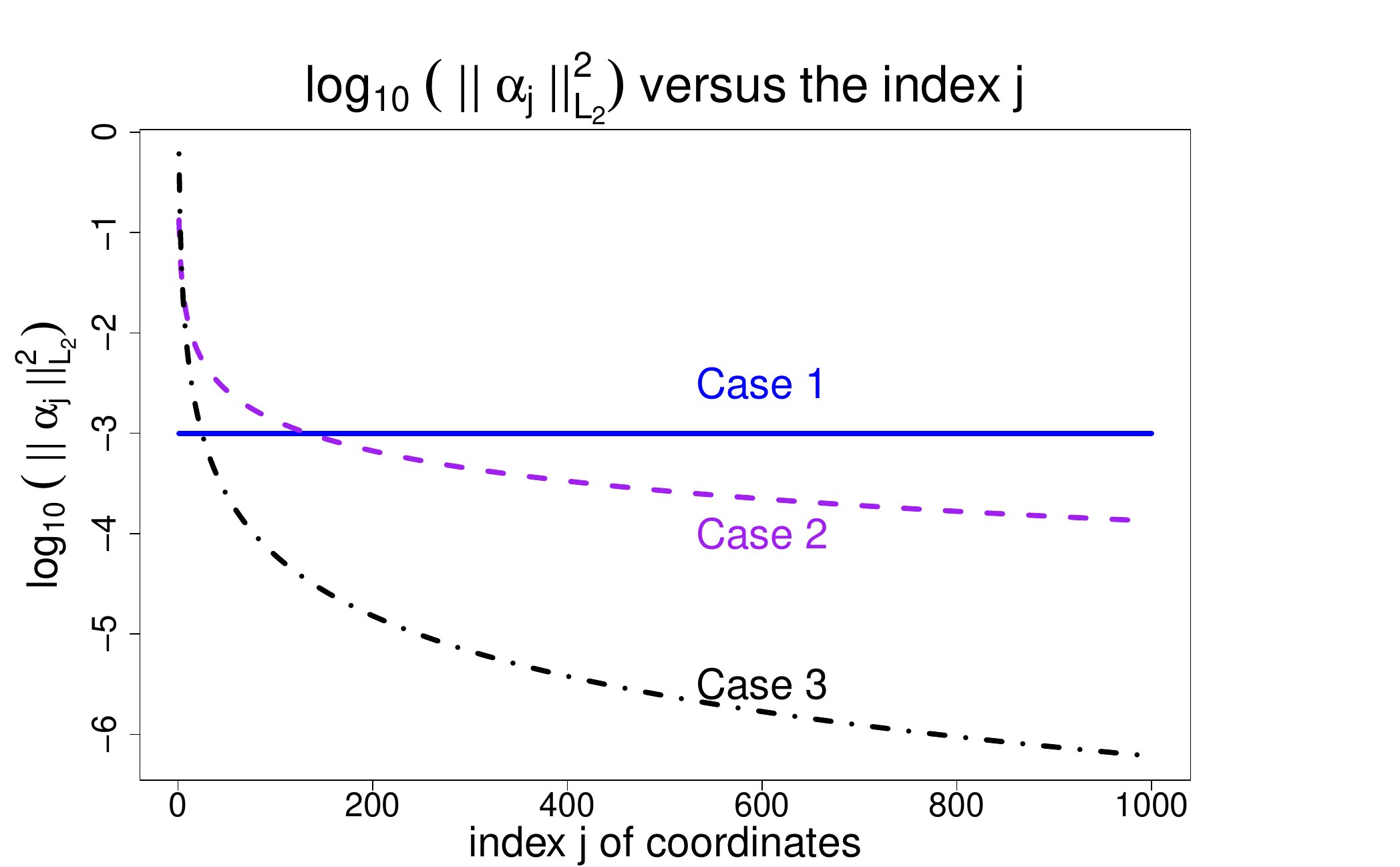}
\end{center}
    \caption{\label{fig_2} \baselineskip=10pt Plots of $\log_{10}(\|\alpha_j\|_{L^2})$ versus $j$ for the three cases: $\|\alpha_j\|_{L^2}^2=0.001, c_1/j, c_2/j^2$, respectively, and $c_1$ and $c_2$ are constants such that $\|\boldsymbol{\alpha}\|_{L^2,2}=1$.
}
\end{figure}

\noindent \underline{\it Situation 2: $p$ is relatively large.} We propose a penalty that simultaneously controls the smoothness and the sparsity of $\boldsymbol{\alpha}(t)$. In Section \ref{sec2.2}, we have shown that  $\widehat{\boldsymbol{\Lambda}}(s,t)$ and $\widehat{\boldsymbol{\Sigma}}(s,t)$ are consistent estimators of $\boldsymbol{\Lambda}(s,t)$ and $\boldsymbol{\Sigma}(s,t)$, respectively, when $p=1$. This result can be extended to the case of a fixed $p>1$ as $n\to\infty$. However, in practice, there are situations where the number $p$ of predictor curves can be much larger than the sample size $n$, such as in neuroscience, the time series curves of thousands of voxels in brain can be obtained. In this case, it is more appropriate to consider the asymptotics as both $p, n\to\infty$.  When both $p, n\to\infty$,  although every entry of $\widehat{\boldsymbol{\Lambda}}(s,t)$ and $\widehat{\boldsymbol{\Sigma}}(s,t)$ converges to the corresponding entry of $\boldsymbol{\Lambda}(s,t)$ and $\boldsymbol{\Sigma}(s,t)$, respectively, for a general $\boldsymbol{\alpha}(t)$, $\int_0^1\int_0^1\boldsymbol{\alpha}(s)\trans\widehat{\boldsymbol{\Lambda}}(s,t)\boldsymbol{\alpha}(t)dsdt$ and $\int_0^1\int_0^1\boldsymbol{\alpha}(s)\trans\widehat{\boldsymbol{\Sigma}}(s,t)\boldsymbol{\alpha}(t)dsdt$ may not converge to $\int_0^1\int_0^1\boldsymbol{\alpha}(s)\trans \boldsymbol{\Lambda}(s,t)\boldsymbol{\alpha}(t)dsdt$ and $\int_0^1\int_0^1\boldsymbol{\alpha}(s)\trans \boldsymbol{\Sigma}(s,t)\boldsymbol{\alpha}(t)dsdt$, respectively. Inspired by the simultaneous EEG-fMRI data, where  the EEG features are only related to the time series curves of a small number of voxels,   we restrict our consideration to  a ``sparse'' $\boldsymbol{\alpha}(t)$. For a scalar vector, a measurement of sparsity is its $l_1$ norm, and sparsity means that most of its coordinates have very small absolute values, which include the case that most coordinates are zero. We extend this concept to a vector of functions. We say $\boldsymbol{\alpha}(t)=(\alpha_1(t),\ldots,\alpha_p(t))\trans$ is sparse if most of the coordinate functions $\alpha_j(t)$ have very small $L^2$ norms. For $\boldsymbol{\alpha}(t)$, we define two norms  
\begin{align}
\|\boldsymbol{\alpha}\|_{L^2,1}=\sum_{j=1}^p\|\alpha_j\|_{L^2} \quad \text{ and } \quad \|\boldsymbol{\alpha}\|_{L^2,2}=\sqrt{\sum_{j=1}^p\|\alpha_j\|_{L^2}^2},\label{56001}
\end{align}
where $\|\boldsymbol{\alpha}\|_{L^2,2}$ is the usual $L^2$-type norm. The $L^1$-type norm $\|\boldsymbol{\alpha}\|_{L^2,1}$ can measure the sparsity of $\boldsymbol{\alpha}(t)$. As an example,  we consider three cases of $\boldsymbol{\alpha}(t)$ with $p=1000$: (1). $\|\alpha_j\|_{L^2}^2=0.001$; (2). $\|\alpha_j\|_{L^2}^2=c_1/j$; (3). $\|\alpha_j\|_{L^2}^2=c_2/j^2$, where $1\le j\le 1000$, and $c_1$ and $c_2$ are constants such that the $L^2$-norms $\|\boldsymbol{\alpha}\|_{L^2,2}$ for the three cases are all equal to one. From Figure \ref{fig_2} which plots $\|\alpha_j\|_{L^2}^2$ versus $j$ for all three cases, we observe that the third case has $\|\alpha_j\|_{L^2}$ decay the fastest and is the sparsest among all three, and the case two is sparser than the case one. The values of $\|\boldsymbol{\alpha}\|_{L^2,1}$ for these three cases are 31.62, 22.59 and 5.83, respectively, which indicates the sparseness of these vectors. 
Arguments in the proof of the asymptotic result in Section \ref{sect.3.3} imply that if $\|\boldsymbol{\alpha}\|_{L^2,1}$ is relatively small compared to $\sqrt{n/\log p}$, $\int_0^1\int_0^1\boldsymbol{\alpha}(s)\trans\widehat{\boldsymbol{\Lambda}}(s,t)\boldsymbol{\alpha}(t)dsdt$ and $\int_0^1\int_0^1\boldsymbol{\alpha}(s)\trans\widehat{\boldsymbol{\Sigma}}(s,t)\boldsymbol{\alpha}(t)dsdt$ converge to $\int_0^1\int_0^1\boldsymbol{\alpha}(s)\trans \boldsymbol{\Lambda}(s,t)\boldsymbol{\alpha}(t)dsdt$ and $\int_0^1\int_0^1\boldsymbol{\alpha}(s)\trans \boldsymbol{\Sigma}(s,t)\boldsymbol{\alpha}(t)dsdt$, respectively. On the other hand, if we assume that $\mathbb{B}(t)$ is sparse (e.g. most of $\mathbf{b}_j(t)$'s in $\mathbb{B}(t)$ are zero), we can show that the $\boldsymbol{\alpha}_k(t)$'s in the optimal decomposition of $\mathbb{B}(t)$ have small $\|\cdot\|_{L^2,1}$ norms, where the sparsity of $\mathbb{B}(t)$ will be defined in the following Section \ref{sect.3.3}. So we first consider to impose a Lasso-type penalty $\tau \|\boldsymbol{\alpha}\|_{L^2,1}^2$ which controls the sparsity of $\boldsymbol{\alpha}(t)$ and ensures the consistency of our estimates.  However, because  $\tau\|\boldsymbol{\alpha}\|_{L^2,1}^2$ is a convex but not strictly convex function of $\boldsymbol{\alpha}$, which leads to computational difficulty in solving the optimization problem, we change $\tau\|\boldsymbol{\alpha}\|_{L^2,1}^2$ to an elasticnet-type penalty $\tau\left[(1-\lambda)\|\boldsymbol{\alpha}\|_{L^2,2}^2+\lambda\|\boldsymbol{\alpha}\|_{L^2,1}^2 \right]$ that is strictly convex. Furthermore, we control the smoothness of the coordinate functions in $\boldsymbol{\alpha}(t)$ by replacing $\|\alpha_j\|_{L^2}^2$ with $\|\alpha_j\|_\eta^2= \|\alpha_j\|_{L^2}^2+\eta\|\alpha^{\prime\prime}_j\|^2_{L^2}$. In summary, we propose the following simultaneous smooth-sparse penalty
\begin{align}
& P(\boldsymbol{\alpha})=\tau\left\{(1-\lambda)\left(\sum_{j=1}^p\|\alpha_j\|^2_\eta\right)+\lambda\left(\sum_{j=1}^p\|\alpha_j\|_\eta\right)^2 \right\},\label{1003}
\end{align}
where $\eta\ge 0$  controls the smoothness,  $0\le \lambda\le 1$ tunes the proportion of the $L_1$ part $\sum_{j=1}^p\|\alpha_j\|_\eta$ and the $L_2$ part $\sum_{j=1}^p\|\alpha_j\|^2_\eta$ in the penalty, and $\tau\ge 0$ controls the strength of the whole penalty. The $\tau$ and $\lambda$ are similar to the tuning parameters in the elasticnet penalty \citep{zou-2005} for high-dimensional data analysis.

In practice, it is to be determined case-by-case regarding whether $p$ is large.  In our empirical studies, we find that there is no significant difference between the two penalties in predictive performances when $p\le 5$, but the method with smooth penalty $\eqref{3677}$ is much faster than the other. When $p>5$, significant improvement of prediction accuracy has been observed using simultaneous smooth-sparse penalty $\eqref{1003}$ in simulation and real data analysis. 

In both of these two situations, after obtaining estimates $\widehat{\boldsymbol{\alpha}}_k(t)$, $1\le k\le K$, as in Section \ref{sec2.2}, we define new scalar predictors: $\widehat{\mathbf{T}}_k=\big(\int_0^1\{\mathbf{X}_1(t)-\overline{\mathbf{X}}(t)\}\trans\widehat{\boldsymbol{\alpha}}_k(t) dt,\ldots, \allowbreak \int_0^1\{\mathbf{X}_n(t)-\overline{\mathbf{X}}(t)\}\trans \widehat{\boldsymbol{\alpha}}_k(t) dt\big)\trans$, and obtain the estimates $\widehat{\boldsymbol{\mu}}$ and $\widehat{\mathbf{w}}_k$'s using the same method as in Section~\ref{sec2}. Finally, we  estimate $\boldsymbol{\mu}_{\mathbf{Y}|\mathbf{X}}$  by  $\widehat{\boldsymbol{\mu}}_{\mathbf{Y}|\mathbf{X}}=\widehat{\boldsymbol{\mu}}+ \sum_{k=1}^K \left[\int_0^1\{\mathbf{X}(t)-\overline{\mathbf{X}}(t)\}\trans\widehat{\boldsymbol{\alpha}}_k(t) dt \right] \widehat{\mathbf{w}}_k$. Given a new observation $\mathbf{X}_{\rm new}(t)$ of the functional predictors, we  predict the response vector as 
\begin{align}
\mathbf{Y}_{\rm pred}=\widehat{\boldsymbol{\mu}}+ \sum_{k=1}^K \left[\int_0^1\{\mathbf{X}_{\rm new}(t)-\overline{\mathbf{X}}(t)\}\trans\widehat{\boldsymbol{\alpha}}_k(t) dt \right]\widehat{\mathbf{w}}_k. 
\label{pred.eq}
\end{align}

\subsection{Asymptotic results}\label{sect.3.3}

We  consider the asymptotic results as both the sample size $n$ and the number $p$ of predictive curves go to infinity, but with the dimension $m$ of the response vector $\mathbf{Y}$ fixed.   Let
\begin{align*}
 \mathbb{Y}=[\mathbf{Y}_1,\ldots, \mathbf{Y}_n]\trans,\quad \mathbb{X}(t)=[\mathbf{X}_1(t),\ldots, \mathbf{X}_n(t)]\trans, \quad \mathbb{E}=[\boldsymbol{\varepsilon}_1,\ldots, \boldsymbol{\varepsilon}_n]\trans
\end{align*}
be the $n\times m$ response matrix, the $n\times p$ matrix of functional predictors and the $n\times m$ noise matrix, respectively. The sample version of model $\eqref{msof.eq}$ can be written as
\begin{align}
 \mathbb{Y}=\mathbb{F}+\mathbb{E}, \quad \text{where} \quad \mathbb{F}=\mathbf{1}_n\boldsymbol{\mu}\trans+\int_0^1\mathbb{X}(t)\mathbb{B}(t)dt \label{3678}
\end{align} 
 is the $n\times m$ data matrix of regression function, and $\mathbf{1}_n$ is the $n$-dimensional vector of 1's. We estimate $\mathbb{F}$ as  
\begin{align}
 \widehat{\mathbb{F}}=\mathbf{1}_n\widehat{\boldsymbol{\mu}}\trans+\sum_{k=1}^K \left[\int_0^1\left\{\mathbb{X}(t)-\mathbf{1}_n\overline{\mathbf{X}}(t)\trans\right\}\widehat{\boldsymbol{\alpha}}_k(t) dt\right] \widehat{\mathbf{w}}_k\trans. \label{3679}
\end{align} 
We will provide a convergence rate for the estimation error $\|\widehat{\mathbb{F}}-\mathbb{F}\|_F$, where $\|\cdot\|_F$ is the Frobenius norm. Moreover, for a new observation $\left(\mathbf{X}_{\rm new}(t), \mathbf{Y}_{\rm new}\right)$, we will provide a convergence rate for the prediction error $\|\mathbf{Y}_{\rm pred}-\mathbf{Y}_{\rm new}\|_2$, where $\mathbf{Y}_{\rm pred}$ is given in $\eqref{pred.eq}$.  We provide two regularity conditions.
\begin{itemize}
\item[(C1).]  For any integer $l\ge 2$,   
\begin{align*}
 \mathrm{E}[\|\boldsymbol{\varepsilon}\|_2^{2l}]\le l!2^{l-1}(\sigma_\epsilon^2)^{l},\quad  \max_{0\le t\le 1}\max_{1\le j\le {p}}\mathrm{E}[X_j(t)^{2l}]\le  l!2^{l-1},
\end{align*} 
where $\sigma_\epsilon^2$ is the sum of the variances of all coordinates of $\boldsymbol{\varepsilon}$.
 \end{itemize}
 The first inequality in (C1) holds for any multivariate normal distribution and $\sigma_\epsilon^2$ is a measure of the noise level.  The second inequality in (C1) holds if  $X_j(t)$ is a Gaussian process with variance not larger than one for any $1\le j\le p$ and $0\le t\le 1$. If the variance of the Gaussian process $X_j(t)$ is greater than one at some time points, we can scale $X_j(t)$ to have variance not larger than one and correspondingly scale $\mathbb{B}(t)$ to keep the model unchanged, and our main results still hold for the scaled $X_j(t)$'s and $\mathbb{B}(t)$. 
For the coefficient matrix $\mathbb{B}(t)=\left[\mathbf{b}_1(t), \ldots, \mathbf{b}_p(t)\right]\trans$, we define an $L_1$ type norm   
\begin{align}
 \|\mathbb{B}\|_{L^2,1}= \sum_{j=1}^p\|\mathbf{b}_j\|_{L^2,2}=\sum_{j=1}^p  \sqrt{\sum_{i=1}^m\|b_{ji}\|_{L^2}^2},
\label{bnorm.eq}
\end{align}
which  measures the sparsity of  $\{\mathbf{b}_1(t),\ldots,\mathbf{b}_p(t)\}$.
\begin{itemize}
\item[(C2).]    $\lim_{n\to\infty}\sqrt{\frac{\log{p}}{n}}=0 \quad\text{and }\quad \lim_{n\to\infty}\left(\sigma_\epsilon^2 +\|\mathbb{B}\|_{L^2,1}^2\right)\sqrt{\frac{\log{p}}{n}}=0$.
  \end{itemize}
This condition implies that $p$ can be much larger than $n$, but should increase slower than $e^n$. Both the noise level $\sigma_\epsilon^2$ and the sparsity  of coefficient functions, $\|\mathbb{B}\|_{L^2,1}^2$, should increase slower than $\sqrt{n/\log{p}}$.  

By Theorem~\ref{theorem_1}, we only need the first few components with relatively large $\sigma_k^2$. Let 
\begin{align}
  K=\max\left\{k: \sigma_k^2 \ge \delta\left(\sigma_\epsilon^2 +\|\mathbb{B}\|_{L^2,1}^2\right)\sqrt{\frac{\log{p}}{n}}\right\}\label{2104}
\end{align} 
for a constant $\delta$ not depending on $n$.

 \begin{Theorem}\label{theorem_2} 
 Assume that conditions (C1) and (C2) hold, and $K$ is chosen based on $\eqref{2104}$ with $\delta$ large enough. Assume that the tuning parameters satisfy
\begin{align}
  \tau=M\sqrt{\frac{\log{p}}{n}}, \quad \lambda_0\le \lambda\le 1,\quad 0\le \eta\le  \min_{1\le k\le K}\{\|\boldsymbol{\alpha}_k\|_{L^2,1}^2/\|\boldsymbol{\alpha}_k^{\prime\prime}\|_{L^2,1}^2\}, \label{112104}
\end{align} 
where $0<\lambda_0<1$ is a constant and $M$ is a constant large enough. Both $\lambda_0$ and $M$ do not depend on $n$. Suppose that  there exists a constant $c_1>0$ not depending on $n$, such that
\begin{align}
   \frac{\sigma_k^2-\sigma_{k+1}^2}{\sigma_k^2} \ge c_1 , \quad \text{for all $n$ and $1\le k\le K$}.\label{12104}
\end{align} 
 Then for any $n\ge 1$,  there exists an event $\Omega_n$ with $P(\Omega_n)\ge 1-12p^{-1}$, such that in $\Omega_n$,  we have
\begin{align}
 &\frac{1}{n}\|\widehat{\mathbb{F}}-\mathbb{F}\|_F^2  \le D_1  \left(\sigma_\epsilon^2 +\|\mathbb{B}\|_{L^2,1}^2\right)\sqrt{\frac{\log{p}}{n}} ,\label{12105}\\
&\mathrm{E}\left[\|\mathbf{Y}_{\rm pred}-\mathbf{Y}_{\rm new}\|_2^2|\mathbb{X},\mathbb{Y}\right]\le D_2  \left(\sigma_\epsilon^2 +\|\mathbb{B}\|_{L^2,1}^2\right)\sqrt{\frac{\log{p}}{n}}, \label{2103}
\end{align} 
where $D_1$ and $D_2$ are constants not depending on $n$.
\end{Theorem}
 
The constant $\lambda_0$ can be any value between 0 and 1. Therefore, the choice of the tuning parameter $\lambda$ is   arbitrary as long as it is bounded away from zero and does not affect the convergence rates of the upper bounds in $\eqref{12105}$ and $\eqref{2103}$. These convergence rates  imply that lower noise levels and sparser functional coefficient matrix can lead to better estimation and prediction accuracy for our method.

\section{Computational issues}\label{section_4} 

To solve the optimization problem $\eqref{12220}$, we represent each function in $\boldsymbol{\alpha}(t)$ using B-spline basis expansion, and transform $\eqref{12220}$ to a generalized eigenvalue problem for the concatenated coefficient vectors of these basis expansions. The number of basis functions does not affect the estimation as long as it is large enough due to the smoothness penalty. Details for solving $\eqref{12220}$ are given in Section S.2 in supplementary material. As in FPLS and FPCA, it is important to choose appropriate number $K$ of components. Different ways of selecting $K$ in FPCA has been proposed, such as hypothesis testing \citep{cardot2003testing, horvath2012inference, su2017hypothesis}, bootstrapping \citep{hall2006assessing}, information criteria \citep{Yao-2005, li2013selecting}, cross-validation \citep{hosseini2013cross, kramer2008penalized, goldsmith2011penalized} and generalize cross-validation \citep{cardot2003spline}. Methods for choosing the tuning parameters include restricted maximum likelihood \citep{wood2011fast}, generalized cross-validation \citep{Wanba-1990, ruppert2003semiparametric, reiss2007functional}, cross-validation \citep{kramer2008penalized, goldsmith2011penalized} and so on. In this paper, we choose the optimal number of components $K_{opt}$ and the optimal tuning parameters simultaneously based on a cross-validation procedure, which is implemented in the \texttt{R} package \texttt{FRegSigCom}. The details are provided in the following section.  
 
\subsection{Choice of the number of components and tuning parameters}\label{sect4.3}
 
Following Section \ref{sec3.2}, we consider the two situations of small $p$ and large $p$ separately.  

\noindent \underline{\it Situation 1: $p$ is small and penalty $\eqref{3677}$ is used.}  There are two tuning parameters, $\tau$ and $\eta$. We choose $\tau$ from the set $\{10^{-9}, 10^{-6}, 10^{-3}, 10^{1}, 10^3\}$ and $\eta$ from $\{10^{-7}, 10^{-5}, 10^{-3}, 10^{-1}, 10\}$. The optimal pair $(\tau_{\rm opt}, \eta_{\rm opt})$ is chosen from the 25 pairs $\{(\tau_i, \eta_j): 1\le i\le 5, 1\le j\le 5\}$, together with the optimal number $K_{\rm opt}$ of components, by a cross-validation procedure, where $\tau_i$'s and $\eta_j$'s are the values in the two candidate sets for $\tau$ and $\eta$, respectively. To provide a range for $K_{\rm opt}$, we notice that Theorem \ref{theorem_1} suggests to select the components with relatively large values of $\sigma_k^2$ which can be estimated by the maximum value $\widehat{\sigma}_k^2$ of the optimization problem $\eqref{12220}$. Moreover, by Theorem \ref{theorem_1}, the number of components cannot exceed the dimension $m$ of the response vector. So we propose an upper bound for $K_{\rm opt}$ as 
 \begin{align}
 & K_{upper}=\min\left\{k>1: \frac{\widehat{\sigma}_k^2}{\widehat{\sigma}_1^2+\cdots+\widehat{\sigma}_k^2} \le 0.001\right\}\bigwedge m, \label{24030}
\end{align} 
where $a\bigwedge b$ denotes the smaller value of $a$ and $b$. The optimal number $K_{\rm opt}$ of components is chosen from $\{1, \ldots, K_{upper}\}$, together with $(\tau_{\rm opt}, \eta_{\rm opt})$ using the following algorithm. 

 \noindent \underline{\it Step 1.} For each pair $(\tau_i, \eta_j)$, $1\le i\le 5$ and $1\le j\le 5$, we calculate the upper bound $K^{(i,j)}_{upper}$  using the whole data set and $\eqref{24030}$.   

 \noindent \underline{\it Step 2.} We randomly split the whole data set into five subsets with roughly equal sizes. For each $ v \in \{ 1, \ldots, 5\}$, let $(\mathbb{Y}^{(v)}, \mathbb{X}^{(v)}(t))$ denote the $v$-th subset which is used as the $v$-th validation set, and all other observations as the $v$-th training set. For each $v \in \{ 1, \ldots, 5\}$ and each candidate pair of tuning parameters $(\tau_i, \eta_j)$,  
\begin{itemize}
\item [(a).] We calculate $\widehat{\boldsymbol{\alpha}}_k^{(v,i, j)}(t)$, $\widehat{\boldsymbol{\mu}}^{(v,i, j)}$ and ${\mathbf{w}}^{(v,i, j)}_k$ for all $k \in \{ 1, \ldots,  K^{(i,j)}_{upper}\}$, using the $v$-th training set.
\item [(b).]
 For each $k \in \{ 1, \ldots, K^{(i,j)}_{upper}\}$, we compute 
\[
\widehat{\mathbb{B}}^{(v,i,j,k)}(t)=\widehat{\boldsymbol{\alpha}}^{(v,i,j)}_1(t) {{\widehat{\mathbf{w}}^{(v,i,j)}_1}{}^T} +\cdots+\widehat{\boldsymbol{\alpha}}^{(v,i,j)}_k(t) {{\widehat{\mathbf{w}}^{(v,i,j)}_k}{}^T}
\]
and the predicted responses $\mathbb{Y}^{(v,i,j,k)}_{\rm pred}=\mathbf{1}_{n_v} {{\widehat{\boldsymbol{\mu}}^{(v,i, j)}}{}^T}+\int_0^1 \mathbb{X}^{(v)}(t) \widehat{\mathbb{B}}^{(v,i,j,k)}(t)dt$ for the $v$-th validation data set,
where  $n_v$ is the number of the observations in the $v$-th validation set. Then we calculate  the  validation error $e_{v,i,j,k}=\|\mathbb{Y}^{(v,i,j,k)}_{\rm pred}-\mathbb{Y}^{(v)}\|_F^2$.
 \end{itemize}
We repeat Steps (a)-(b) for all $1\le v\le 5$, $1\le i\le 5$ and $1\le j\le 5$.

\noindent \underline{\it Step 3.} We compute the total validation error, ${e}^{(total)}_{i, j, k}=\sum_{v=1}^5 e_{v,i,j, k}$, for all $i, j, k$.   Let ${e}^{(total)}_{i_0, j_0, k_0}=\min\{{e}^{(total)}_{i,j,k}: 1\le i\le 5, 1\le j\le 5,  1\le k\le \widehat{K}^{(i,j)}_{upper}\}$. Our optimal tuning parameters are $\tau_{\rm opt}=\tau_{i_0}$, $\eta_{\rm opt}=\eta_{j_0}$, and  $\widehat{K}_{\rm opt}=k_0$. \\
  
\noindent \underline{\it Situation 2: $p$ is large, and penalty $\eqref{1003}$ is used}. We use the same cross-validation procedure as above except that there are three tuning parameters, $\tau$, $\lambda$ and $\eta$, in penalty $\eqref{1003}$. In general, people choose the three tuning parameters in a three-dimensional grid to explore various combinations of them. However, as the penalty $\eqref{1003}$  involves the $L^1$ type norm and may be applied to data with hundreds or thousands predictor curves, it will be much more complicated and time-consuming than the first situation. To improve the computational efficiency of our algorithm for a large $p$, we reduce the number of candidate values for the triple parameters. In penalty $\eqref{1003}$, the term $\left(\sum_{j=1}^p\|\alpha_j\|_\eta\right)^2$ plays the main role for curve selection and is tuned by $\tau\lambda$. So instead of exploring all possible pairs of $(\tau, \lambda)$ in a two-dimensional grid, we select them from the set $\{(10^{-1},0.1), (1,0.2), (10,0.3), (100,0.4)\}$, where the values of the multiplication $\tau\lambda$ spread over a wide range. The smoothness tuning parameter $\eta$ is chosen from $\{10^{-6},10^{-3},1\}$. So the optimal triple $(\tau_{\rm opt}, \lambda_{\rm opt}, \eta_{\rm opt})$ is chosen from $4\times 3=12$ triple values.

 \section{Simulation studies}\label{section_5}
We conduct simulation studies for $p=1$ and $p>1$ separately. For each simulation, the intercept $\boldsymbol{\mu}$ is an $m$-dimensional vector with all coordinates equal to one, and the noise vector $\boldsymbol{\varepsilon}=(\varepsilon_1,\ldots, \varepsilon_m)$, where $\varepsilon_k$'s are independent random variables from the same distribution $N(0, \sigma^2)$. We consider three noise levels, $\sigma=0.01, 0.1, 1$ and three values of dimension $m$, $m=1, 5, 10$. Because the wavelet-based methods conduct the fast wavelet transformation which requires the number of the observation time points to be a power of two, in all the simulations, the predictive curves are all observed at 64 equally spaced points in $[0,1]$.

\subsection{Simulations with one predictor curve}\label{sec5.1}
We compare our method (\textit{new.Sm}, implemented in the \texttt{R} package \texttt{FRegSigCom}) to the following methods. The penalized functional regression (\textit{pfr}, \citet{goldsmith2011penalized}) expands the sample predictor curves using the principal component basis and the coefficient function using spline basis, and is implemented in the \texttt{R} package \texttt{refund}. The functional regression using basis representation (\textit{fbasis}, \citet{Ramsay-Silverman-2005} and \citet{febrero2012statistical}) expands both the predictor curves and the coefficient function using spline basis, and is implemented in the \texttt{R} package \texttt{fda.usc}. The functional regression using principal components analysis (\textit{fpcr}, \citet{cardot1999functional}) expands both the predictor curves and the unknown coefficient function using the principal component basis, and is implemented in the \texttt{R} package \texttt{fda.usc}. The functional penalized PLS regression with scalar response (\textit{fpls}, \citet{preda2005pls} and \citet{kramer2008penalized}) expands both the predictor curves and the unknown coefficient function using the FPLS  basis, and is implemented in the \texttt{R} package \texttt{fda.usc}. The  functional regression using nonparametric kernel estimation (\textit{fnp}, \citet{ferraty2006nonparametric}) is implemented in the \texttt{R} package \texttt{fda.usc}. The  Gaussian process regression for multivariate scalar-on-function regression
 (\textit{gp-msof}, \citet{wang2017gaussian}) conducts PCA on the multivariate response and then fits a nonparametric model for each PC score using the Gaussian process regression \citep{o1978curve, rasmussen2006gaussian}. The code for \textit{gp-msof} is provided by the authors of \citet{wang2017gaussian}. 
Two wavelet based methods, \textit{wLasso} \citep{zhao2012wavelet} and \textit{wwLasso} \citep{zhao2015wavelet}, first transform the predictor curves to the wavelet coefficient vectors using the fast wavelet transformation, and then apply Lasso or weighted Lasso to the wavelet coefficients. We use the \texttt{R} package \texttt{wavethresh} to perform the fast wavelet transformation and the \texttt{R} package \texttt{glmnet} to perform Lasso and weighted Lasso. 
In \textit{new.Sm}, \textit{pfr} and \textit{fbasis}, the numbers of B-spline basis functions is chosen to be the same and equal to 30 for all simulations and real data analysis. For \textit{fbasis}, \textit{fpcr} and \textit{fpls}, we choose the smooth tuning parameter from the set $\{10^{-10}, 10^{-9}, 10^{-8}, \ldots, 1, 10, \ldots, 10^9, 10^{10}\}$, which leads to better performance than the default settings. In \textit{wLasso},  we choose the option \texttt{family="mgaussian"} in function \texttt{cv.glmnet} to jointly fit the multivariate response by Lasso. However, in \textit{wwLasso}, if we choose the option \texttt{family="mgaussian"}, the prediction error will greatly increase, so we fit each coordinate of the response separately. 

 We perform two sets of simulations with different ways to generate the predictor curves and coefficient functions.  

\noindent\underline{\large\it Simulation 1.}   
$X(t)$ is generated from a Gaussian process with covariance function $\text{exp}\{-[30(t-t')]^2\}$. The coefficient $\mathbf{b}(t)=(b_1(t),\ldots, b_m(t))\trans$ is given by $b_k(t)=\sin(3\pi t+k)+\cos(3\pi t+k)$, $1\le k\le m$.  

\noindent\underline{\large\it Simulation 2.}   
 $X(t)$ is generated by the basis expansion, $X(t)=\sum_{k=1}^{20}[V_{k1}\sin(2k\pi t)+V_{k2}\cos(2k\pi t)]$, where $V_{kj}\sim N(0, 1/k^{1/2})$ for $1\le k\le 20$ and $j=1,2$ are independent.  The coefficient $\mathbf{b}(t)=\mathbb{M}\boldsymbol{\beta}(t)$, where $\mathbb{M}$ is an $m\times 3$ matrix with each entry independently generated from the uniform distribution in $(0,1)$,  $\boldsymbol{\beta}(t)=(\beta_1(t),\beta_2(t), \beta_3(t))\trans$, with $\beta_1(t)=\cos(2\pi t)$, $\beta_2(t)=2t^2$ and $\beta_1(t)=1/(1+t)$.

Sample curves of $X(t)$ in these two simulations are plotted in Figure S.1  of supplementary material. We generate data from the model $\mathbf{Y}=\boldsymbol{\mu} +c\int_0^1X(t)\mathbf{b}(t) dt+\boldsymbol{\varepsilon}$, where $c$ is a scaling constant such that when the noise level $\sigma=1$, the signal-to-noise ratio is equal to 1. For each setting, we repeat the following procedure 100 times. In each repeat, we generate a training data of size 100 and a test data set  $\{X_{{\rm test}, \ell}(t), \mathbf{Y}_{{\rm test},\ell}: 1\le \ell\le 500\}$ of size 500. We use the training set to select tuning parameters and fit model, and apply the fitted model to the test data to calculate the mean squared prediction error (MSPE)
\begin{align*}
  {\rm MSPE}=\frac{1}{500}\frac{\sum_{\ell=1}^{500}\|\widehat{\mathbf{Y}}_\ell^{\rm pred}-\mathbf{F}_{{\rm test},\ell}\|_2^2}{m},
\end{align*}
for each method, where $\widehat{\mathbf{Y}}_\ell^{\rm pred}$ is the predicted response vector for $X_{{\rm test}, \ell}(t)$ and $\mathbf{F}_{{\rm test},\ell}=\boldsymbol{\mu} +\int_0^1X_{{\rm test},\ell}(t)\mathbf{b}(t) dt$. 
 The averages and standard deviations of the MSPEs from 100 replicates for Simulations 1 and 2 are summarized in Table~\ref{sim1.tab} which show some interesting patterns.

\noindent (1). When the noise level is relatively small ($\sigma=0.01, 0.1$), our method has a large advantage over other methods, especially when $m$ is relatively large. For example, when $m=10$ and $\sigma=0.01$, the averaged MSPEs of other methods are at least 24 times higher than that of our method in Simulation 1, and 6.3 times higher in Simulation 2.

\noindent (2). When the noise level is relatively large ($\sigma=1$), for $m=1$, our method, \textit{pfr} and \textit{fbasis} have close MSPEs which are lower than others in both simulations; when $m>1$, our method has the best predictive performance. 

\noindent (3). The MSPEs of our method decrease when $m$ increases in both simulations for all noise levels. In Simulation 2, for any $m$, the coordinates of the response have the same distribution which is also the same as that for $m=1$. So when $m>1$, if we separately fit the model for each coordinate of the response, the MSPEs should be the same as those of $m=1$. By taking care of the correlation structure in response, our method has improved predictive accuracy when $m$ is larger. In Simulation 2, for $\sigma=0.1$, the averaged number of components chosen by our method is 3.34 for $m=5$ and 3.26 for $m=10$. Compared to the case of $m=5$, despite there are more coordinates when $m=10$, our method chooses less components. So our method makes more aggressive dimension reduction when $m=10$ than when $m=5$, which leads to improvement of prediction accuracy. Similarly we can explain the pattern in Simulation 1.

\begin{table}[H]
\caption{\baselineskip=10pt The averages and standard deviations (in parenthesis) of MSPEs  in Simulations 1 and 2.}
\label{sim1.tab}\centering
\renewcommand{\arraystretch}{0.65}
\begin{tabular}{|c|c|c|c|c|c|c|}
\hline\hline
\multicolumn{7}{|c|}{\bf Simulation 1}\\ 
\hline
  $\sigma$ & $m$ &\textit{new.Sm}  & \textit{pfr} & \textit{fbasis} & \textit{fpcr}  & \textit{fpls}   \\\hline
    \multirow{3}{*}{0.01} & 1  &
0.28(0.11)$\times 10^{-4}$ & 5.3(3.8)$\times 10^{-4}$ & 3.6(0.9)$\times 10^{-4}$ & 0.34(0.16) & 0.002(0.0003)  \\
 &5&
0.14(0.04)$\times 10^{-4}$ & 4.2(1.6)$\times 10^{-4}$ & 2.3(0.4)$\times 10^{-4}$ & 0.30(0.10) & 0.001(0.0002)  \\
 &10&
0.09(0.02)$\times 10^{-4}$ & 4.0(1.4)$\times 10^{-4}$ & 2.2(0.4)$\times 10^{-4}$ & 0.31(0.11) & 0.001(0.0002)  \\\hline
\multirow{3}{*}{0.1} & 1  &
1.8(0.7)$\times 10^{-3}$ & 4.9(0.8)$\times 10^{-3}$ & 4.0(1.0)$\times 10^{-3}$ & 0.37(0.16) & 2.8(0.6)$\times 10^{-3}$  \\
 &5&
0.9(0.3)$\times 10^{-3}$ & 3.4(0.4)$\times 10^{-3}$ & 3.0(0.4)$\times 10^{-3}$ & 0.30(0.11) & 2.2(0.3)$\times 10^{-3}$  \\
 &10&
0.6(0.1)$\times 10^{-3}$ & 3.4(0.3)$\times 10^{-3}$ & 3.0(0.3)$\times 10^{-3}$ & 0.31(0.12) & 2.1(0.3)$\times 10^{-3}$  \\\hline
  \multirow{3}{*}{1} & 1  &
0.10(0.05) & 0.09(0.04) & 0.10(0.05) & 0.49(0.20) & 0.23(0.08)\\
 &5&
0.07(0.03) & 0.08(0.02) & 0.10(0.03) & 0.43(0.11) & 0.20(0.04)\\
 &10&
0.05(0.02) & 0.09(0.02) & 0.10(0.02) & 0.46(0.11) & 0.21(0.04) 
\\\hline
$\sigma$ & $m$ &  \textit{fnp} & \textit{gp-msof} & \textit{wLasso} & \textit{wwLasso} & \\\hline
    \multirow{3}{*}{0.01} & 1  &
0.90(0.04) & 0.99(0.03) &  0.002(0.0003)    & 0.002(0.0004) &    \\
 &5&
0.89(0.03) & 1.02(0.02) &  0.002(0.0002)    & 0.002(0.0003) &    \\
 &10&
0.90(0.03) & 1.02(0.02) &  0.002(0.0002)    & 0.001(0.0002) &    \\
\hline
\multirow{3}{*}{0.1} & 1  &
0.91(0.04) & 1.02(0.02) &  6.3(1.9)$\times 10^{-3}$   & 5.1(1.3)$\times 10^{-3}$ &    \\
 &5&
0.90(0.03) & 1.02(0.02) &  4.1(0.8)$\times 10^{-3}$  & 4.7(0.5)$\times 10^{-3}$ &    \\ &10&
0.90(0.03) & 1.02(0.02) &  3.6(0.5)$\times 10^{-3}$   &4.6(0.4)$\times 10^{-3}$ &    \\\hline
  \multirow{3}{*}{1} & 1  &
0.92(0.05) & 1.00(0.03) &  0.39(0.12)   & 0.32(0.09) &    \\
 &5&
0.92(0.03) & 1.02(0.02) &  0.25(0.05)   & 0.31(0.04) &    \\
 &10&
0.92(0.03) & 1.02(0.02) &  0.21(0.04)   & 0.31(0.02) &    \\
\hline\hline
\multicolumn{7}{|c|}{\bf Simulation 2}\\ 
\hline
  $\sigma$ & $m$ &\textit{new.Sm}  & \textit{pfr} & \textit{fbasis} & \textit{fpcr}  & \textit{fpls}   \\\hline
    \multirow{3}{*}{0.01} & 1  &
1.8(0.7)$\times 10^{-5}$ & 2.3(2.7)$\times 10^{-3}$ & 6.8(2.7)$\times 10^{-5}$ & 0.13(0.08) & 0.014(0.011)  \\
 &5&
1.4(0.3)$\times 10^{-5}$ & 1.9(0.5)$\times 10^{-3}$ & 6.3(0.9)$\times 10^{-5}$ & 0.12(0.05) & 0.011(0.002)  \\
 &10&
1.0(0.2)$\times 10^{-5}$ & 1.8(0.4)$\times 10^{-3}$ & 6.3(0.6)$\times 10^{-5}$ & 0.11(0.05) & 0.011(0.002)  \\\hline
\multirow{3}{*}{0.1} & 1  &
1.3(0.6)$\times 10^{-3}$ & 6.9(0.4)$\times 10^{-3}$ & 3.7(1.2)$\times 10^{-3}$ & 0.13(0.07) & 0.015(0.01)  \\
 &5&
0.9(0.2)$\times 10^{-3}$ & 5.8(0.5)$\times 10^{-3}$ & 3.5(0.5)$\times 10^{-3}$ & 0.12(0.05) & 0.013(0.003)  \\
 &10&
0.8(0.2)$\times 10^{-3}$ & 5.8(0.8)$\times 10^{-3}$ & 3.6(0.4)$\times 10^{-3}$ & 0.12(0.06) & 0.013(0.002)  \\\hline
  \multirow{3}{*}{1} & 1  &
0.06(0.04) & 0.06(0.03) & 0.08(0.05) & 0.59(0.24) & 0.15(0.05)\\
 &5&
0.05(0.02) & 0.06(0.02) & 0.08(0.03) & 0.59(0.13) & 0.15(0.04)\\
 &10&
0.04(0.01) & 0.06(0.01) & 0.08(0.02) & 0.59(0.11) & 0.16(0.03) 
\\\hline
$\sigma$ & $m$ &  \textit{fnp} & \textit{gp-msof} & \textit{wLasso} & \textit{wwLasso} & \\\hline
    \multirow{3}{*}{0.01} & 1  &
0.91(0.04) & 1.01(0.03) &  0.0014(0.0003)    & 0.0015(0.0003) &    \\
 &5&
0.90(0.03) & 1.01(0.02) &  0.0014(0.0002)    & 0.0015(0.0003) &    \\
 &10&
0.90(0.03) & 1.00(0.02) &  0.0015(0.0003)    & 0.0016(0.0003) &    \\
\hline
\multirow{3}{*}{0.1} & 1  &
0.90(0.04) & 1.00(0.03) &  4.7(1.4)$\times 10^{-3}$   & 4.4(1.4)$\times 10^{-3}$ &    \\
 &5&
0.90(0.04) & 1.01(0.03) &  3.3(0.7)$\times 10^{-3}$  & 4.0(0.6)$\times 10^{-3}$ &    \\ &10&
0.90(0.03) & 1.02(0.01) &  3.1(0.5)$\times 10^{-3}$   & 4.2(0.4)$\times 10^{-3}$ &    \\\hline
  \multirow{3}{*}{1} & 1  &
0.91(0.04) & 1.01(0.02) &  0.29(0.06)   & 0.22(0.05) &    \\
 &5&
0.91(0.03) & 1.02(0.03) &  0.16(0.04)   & 0.20(0.03) &    \\
 &10&
0.92(0.04) & 1.02(0.03) &  0.14(0.03)   & 0.21(0.02) &    \\
\hline
\end{tabular}
\end{table}

\noindent (4). Our method, \textit{fpls} and \textit{fpca} all consider decompositions of the coefficient functions. The \textit{fpcr} only uses $X(t)$ to construct the component functions and ignores information in $\mathbf{Y}$. The \textit{fpls} and our method use information in both $X(t)$ and $\mathbf{Y}$, but our method estimates the optimal decomposition for prediction. So our method outperforms \textit{fpls}, which is better than \textit{fpca}, in these simulations.

\noindent (5). When $\sigma=0.1$, the two wavelet based methods have similar performance with \textit{pfr} and \textit{fbasis}. As $m$ increases, \textit{wLasso} has decreased prediction error as the multivariate response is jointly fitted and the correlation structure in response has been taken into account.

\noindent (6). All the data in these simulations are generated from linear model. But \textit{fnp} is a nonparametric method which is not specialized for linear model. The high MSPEs of \textit{fnp} may be due to its complication. So does the \textit{gp-msof} which also uses nonparametric method. 

\subsection{Simulations with multiple predictor curves}\label{sec5.2}
When $p$ is moderate or large, many methods in Simulations 1 and 2 are not computationally feasible. We compare our methods, {\it new.SmSp} with the simultaneous smooth-sparse penalty in $\eqref{1003}$ and \textit{new.Sm} with the smooth only penalty in $\eqref{3677}$, with  \textit{wLasso}  and \textit{wwLasso}. In addition, we consider two more hybrid methods \textit{FPCA+gSCAD} \citep{lian2013shrinkage, kong2016partially} and \textit{FPCA+gLasso} \citep{zhu2009functional}, which perform FPCA on each functional predictor using the \texttt{R} package \texttt{fdapace}, and then use the PC scores as new scalar predictors to fit penalized regression model with the \texttt{R} package \texttt{grpreg} by imposing the group SCAD or group Lasso  penalty, respectively. The number of components in FPCA is chosen by the criterion of 99\% variations explained. But these two FPCA based methods are not computationally feasible for large $p$, such as 1000.

 We generate the predictor curves and coefficient functions in the following two different ways.

\noindent\underline{\large\it Simulation 3.}  We take $p=50$.\\
\noindent (1). $\mathbf{X}(t)=(X_1(t),\ldots, X_{50}(t))\trans$ is generated by  basis expansion 
$$X_j(t)=\sum_{k=1}^{20}[V^{(j)}_{k1}\sin(2k\pi t)+V^{(j)}_{k2}\cos(2k\pi t)], \quad 1\le j\le p,$$
 where $\mathbf{V}_{ki}=(V^{(1)}_{ki}, \ldots, V^{(50)}_{ki})\trans$ has a multivariate normal distribution with mean zero and covariance matrix with diagonal entries equal to $1/\sqrt{k}$ and off-diagonal entries  equal to $\rho/\sqrt{k}$, for $1\le k\le 20$ and $i=1,2$. The $\mathbf{V}_{ki}$'s are independent. The value $0\le \rho\le 1$ controls the correlation between $X_j(t)$'s. The larger is $\rho$, the stronger is the correlation. We consider two values $\rho=0.2, 0.7$.\\
 \noindent (2). The coefficient $\mathbb{B}(t)=\left[\boldsymbol{\beta}_1(t), \boldsymbol{\beta}_2(t), \boldsymbol{\beta}_3(t)\right]\mathbb{M}$ is a $p\times m$ matrix, where $\mathbb{M}$ is a $3\times m$ matrix with each entry independently generated from the uniform distribution in $(0,1)$. For $1\le k\le 3$, $\boldsymbol{\beta}_k(t)=(\beta_{1k}(t),\ldots, \beta_{pk}(t))\trans$ is a $p$-dimensional vector with  $\beta_{1k}(t)=t^k$, $\beta_{2k}(t)=\cos(k\pi t)$, $\beta_{3k}(t)=1/(t+k)$, $\beta_{4k}(t)=\log(t+k)$, $\beta_{5k}(t)=e^{-kt^2}$, and $\beta_{jk}(t)=0$ if $j>5$. So only the first five predictor curves have nonzero coefficients and are viewed as true features. 

\noindent\underline{\large\it Simulation 4.}    We take $p=1000$.\\
\noindent (1). To generate $\mathbf{X}(t)=(X_1(t),\ldots, X_{1000}(t))\trans$, we first generate independent curves $W_j(t)$'s from a Gaussian processes with covariance function $\text{exp}\{-[30(t-t')]^2\}$, $1\le j\le 1000+Lag$, and let $X_j(t)=[W_{j+1}(t)+\cdots+W_{j+Lag}(t)]/\sqrt{Lag}$, where $Lag$ is a positive integer used to control the correlation between $X_j(t)$'s. A larger $Lag$ leads to stronger correlation between $X_j(t)$'s. We take $Lag=2, 5$.\\
 \noindent (2). The coefficient $\mathbb{B}(t)=\left[\boldsymbol{\beta}_1(t), \boldsymbol{\beta}_2(t), \boldsymbol{\beta}_3(t)\right]\mathbb{M}$ is a $p\times m$ matrix, where $\mathbb{M}$ is defined in the same way as in Simulation 3. 
For $1\le k\le 3$, $\boldsymbol{\beta}_k(t)=(\beta_{1k}(t),\ldots, \beta_{pk}(t))\trans$ is a $p$-dimensional vector with  $\beta_{1,k}(t)=2(t+1)e^{-kt}$, $\beta_{11, k}(t)=\sin(k\pi t/2)/\sqrt{1+t}$, $\beta_{21, k}(t)=3\sinh(-t)/k+t^2$, $\beta_{31, k}(t)=0.5(1+t)^k \cos(k\pi t)$, $\beta_{41, k}(t)=\tan(t)/(1+kt^2)$ and $\beta_{jk}(t)=0$ for all other $j$'s. So only the five predictor curves, $X_1(t), X_{11}(t), X_{21}(t),X_{31}(t), X_{41}(t)$, have nonzero coefficients and are viewed as true features. 
 
 We  generate data from model $\mathbf{Y}_\ell=\mathbf{F}_\ell+\boldsymbol{\varepsilon}_\ell$, where $\mathbf{F}_\ell=\boldsymbol{\mu} +c\int_0^1\mathbf{X}_\ell(t)\trans\mathbb{B}(t) dt$, $1\le \ell\le 100$, and $c$ is a scaling factor such that when the noise level $\sigma=1$, the signal-to-noise ratio is equal to 1. We conduct similar procedure as in Section \ref{sec5.1} by performing 100 replications, where each replicate has training data of size 100 and test data of size 500.  The averages and standard deviations of the MSPEs from 100 replicates are summarized in Table~\ref{sim3.tab}. In both simulations, our method \textit{new.SmSp} has the lowest MSPEs. The introduction of the simultaneous smooth-sparse penalty greatly improves the performance of our method compared to that without sparsity penalty. With the increase of $m$, the MSPEs of our method \textit{new.SmSp} decreases, especially when $\sigma=1$.

   In addition, we calculate the sensitivity and specificity of each method for curve selection.  Let $\widehat{\mathbb{B}}(t)=\left[\widehat{\mathbf{b}}_1(t), \ldots, \widehat{\mathbf{b}}_p(t)\right]\trans$ be the estimate of the coefficient $\mathbb{B}(t)=\left[\mathbf{b}_1(t), \ldots, \mathbf{b}_p(t)\right]\trans$, where $\mathbf{b}_j(t)$ is the coefficient of the $j$-th predictor curve, $1\le j\le p$. Sensitivity and specificity are respectively defined as
\[\text{sensitivity}=\frac{\text{the number of $j$'s with both $\mathbf{b}_j(t) \neq \mathbf{0}$ and $\widehat{\mathbf{b}}_j(t) \neq \mathbf{0}$}}{\text{the number of $j$'s with $\mathbf{b}_j(t) \neq \mathbf{0}$}}\]
and
\[\text{specificity}=\frac{\text{the number of $j$'s with both $\mathbf{b}_j(t)=\mathbf{0}$ and $\widehat{\mathbf{b}}_j(t)=\mathbf{0}$}}{\text{the number of $j$'s with $\mathbf{b}_j(t)=\mathbf{0}$}}.\]
 Tables S.1 and S.2  of supplementary material summarize the sensitivity and specificity of each method. The method \textit{new.Sm} has sensitivity equal 1 and specificity equal 0 because it does not do variable selection and includes all variables in model fitting. 
Our method \textit{new.SmSp} has the highest sensitivity and  lower specificity than wavelet-based methods or FPC-based methods. That implies that {\it new.SmSp} selects more curves than other methods, including both true feature curves and non-feature curves.

\section{Applications}\label{section_6}
We consider two data sets: the corn data with single predictor curve  and the simultaneous EEG-fMRI data with a large number of predictor curves.

\subsection{Corn data}\label{section_6.1}
This data set consists of 80 samples of corn, and is available at Eigenvector Research (http://www.eigenvector.com/data/Corn/index.html). The moisture, oil, protein and starch values of each corn sample are measured together with its NIR spectrum curve with wavelength range from 1100nm to 2498nm  at every 2 nm. 

\begin{table}[H]
\caption{\baselineskip=10pt The averages (and standard deviations) of MSPEs  in Simulations 3 and 4.}
\label{sim3.tab}\centering
\small\addtolength{\tabcolsep}{-3pt}
 \renewcommand{\arraystretch}{0.75}
 {
\begin{tabular}{|c|c|c|c|c|c|c|c|c|}
\hline
  \multicolumn{9}{|c|}{\bf Simulation 3}\\\hline
$\rho$ & $\sigma$ & $m$ &\textit{new.SmSp} & \textit{new.Sm}   & \textit{wLasso}   &\textit{wwLasso} & \textit{FPCA+gSCAD} & \textit{FPCA+gLasso} \\\hline\hline
\multirow{9}{*}{0.2} &\multirow{3}{*}{0.01} & 1  &
 0.050(0.026) & 0.349(0.165) & 0.375(0.156) & 0.186(0.073)  & 0.605(0.135) & 0.628(0.121) \\
&&5&
 0.049(0.015) & 0.364(0.093) & 0.308(0.094) & 0.184(0.051)  & 0.602(0.099) & 0.625(0.096) \\
&&10&
 0.049(0.013) & 0.360(0.075) & 0.300(0.088) & 0.181(0.046)  & 0.599(0.083) & 0.624(0.085) \\\cline{2-9}
&\multirow{3}{*}{0.1} & 1  &
 0.064(0.030) & 0.363(0.174) & 0.379(0.156) & 0.198(0.088)  & 0.598(0.133) & 0.628(0.125) \\
&&5&
 0.057(0.017) & 0.360(0.082) & 0.316(0.092) & 0.195(0.048)  & 0.610(0.098) & 0.637(0.097) \\
&&10&
 0.053(0.013) & 0.364(0.084) & 0.317(0.086) & 0.197(0.042)  & 0.611(0.088) & 0.628(0.093) \\\cline{2-9}
&\multirow{3}{*}{1} & 1  &
 0.536(0.200) & 0.883(0.151) & 0.893(0.155) & 0.724(0.205)  & 0.952(0.108) & 0.957(0.114) \\
&&5&
 0.267(0.068) & 0.687(0.104) & 0.667(0.125) & 0.679(0.108)  & 0.940(0.080) & 0.940(0.082) \\
&&10&
 0.228(0.045) & 0.593(0.093) & 0.603(0.109) & 0.677(0.104)  & 0.941(0.060) & 0.943(0.061) \\\hline
\multirow{9}{*}{0.7} &\multirow{3}{*}{0.01} & 1  &
 0.051(0.022) & 0.174(0.076) & 0.240(0.096) & 0.113(0.052)  & 0.360(0.103) & 0.349(0.088) \\
&&5&
 0.049(0.013) & 0.180(0.043) & 0.204(0.056) & 0.106(0.028)  & 0.359(0.065) & 0.347(0.056) \\
&&10&
 0.047(0.011) & 0.182(0.038) & 0.207(0.045) & 0.107(0.028)  & 0.357(0.059) & 0.349(0.053) \\\cline{2-9}
&\multirow{3}{*}{0.1} & 1  &
 0.059(0.023) & 0.191(0.081) & 0.256(0.100) & 0.125(0.054)  & 0.361(0.089) & 0.357(0.086) \\
&&5&
 0.054(0.013) & 0.188(0.046) & 0.216(0.051) & 0.121(0.033)  & 0.368(0.059) & 0.356(0.054) \\
&&10&
 0.051(0.012) & 0.184(0.035) & 0.213(0.044) & 0.121(0.030)  & 0.363(0.055) & 0.354(0.051) \\\cline{2-9}
&\multirow{3}{*}{1} & 1  &
 0.436(0.153) & 0.537(0.172) & 0.670(0.192) & 0.574(0.193)  & 0.733(0.142) & 0.736(0.151) \\
&&5&
 0.276(0.064) & 0.445(0.072) & 0.435(0.085) & 0.537(0.098)  & 0.709(0.075) & 0.707(0.072) \\
&&10&
 0.230(0.042) & 0.377(0.059) & 0.385(0.063) & 0.522(0.077)  & 0.699(0.061) & 0.695(0.059) \\\hline
\end{tabular}
\begin{tabular}{|c|c|c|c|c|c|c|}
\hline
  \multicolumn{7}{|c|}{\bf Simulation 4}\\\hline
$Lag$ & $\sigma$ & $m$ &\textit{new.SmSp} & \textit{new.Sm}   & \textit{wLasso}   &\textit{wwLasso}   \\\hline\hline
\multirow{9}{*}{2} &\multirow{3}{*}{0.01} & 1  &
 0.187(0.109) & 0.973(0.043) & 0.465(0.143) & 0.378(0.138)    \\
&&5&
 0.174(0.066) & 0.968(0.040) & 0.394(0.098) & 0.351(0.097)  \\
&&10&
 0.148(0.042) & 0.971(0.040) & 0.372(0.073) & 0.350(0.077)   \\\cline{2-7}
&\multirow{3}{*}{0.1} & 1  &
 0.207(0.120) & 0.976(0.040) & 0.465(0.154) & 0.382(0.144)    \\
&&5&
 0.163(0.062) & 0.971(0.041) & 0.399(0.117) & 0.376(0.110)    \\
&&10&
 0.171(0.052) & 0.974(0.040) & 0.388(0.091) & 0.370(0.091)   \\\cline{2-7}
&\multirow{3}{*}{1} & 1  &
 0.657(0.244) & 1.03(0.06) & 0.884(0.160) & 0.847(0.184)  \\
&&5&
 0.328(0.086) & 0.997(0.043) & 0.632(0.131) & 0.785(0.124)  \\
&&10&
 0.274(0.063) & 0.987(0.041) & 0.572(0.095) & 0.775(0.119)  \\\hline
\multirow{9}{*}{5} &\multirow{3}{*}{0.01} & 1  &
 0.181(0.132) & 0.953(0.052) & 0.459(0.184) & 0.363(0.168)  \\
&&5&
 0.154(0.050) & 0.843(0.052) & 0.376(0.088) & 0.330(0.089)   \\
&&10&
 0.149(0.053) & 0.944(0.050) & 0.362(0.099) & 0.318(0.094)   \\\cline{2-7}
&\multirow{3}{*}{0.1} & 1  &
 0.180(0.106) & 0.945(0.052) & 0.437(0.155) & 0.348(0.138)   \\
&&5&
 0.164(0.059) & 0.947(0.053) & 0.392(0.100) & 0.353(0.098)   \\
&&10&
 0.168(0.058) & 0.954(0.055) & 0.379(0.088) & 0.352(0.088)  \\\cline{2-7}
&\multirow{3}{*}{1} & 1  &
 0.645(0.220) & 1.02(0.054) & 0.901(0.137) & 0.848(0.166)   \\
&&5&
 0.334(0.112) & 0.987(0.056) & 0.635(0.143) & 0.772(0.130)   \\
&&10&
 0.259(0.051) & 0.958(0.054) & 0.553(0.090) & 0.737(0.114)  \\\hline
\end{tabular}}
\end{table}

\begin{figure}[h]
\begin{center}
\includegraphics[height=2.5in,width=6.5in]{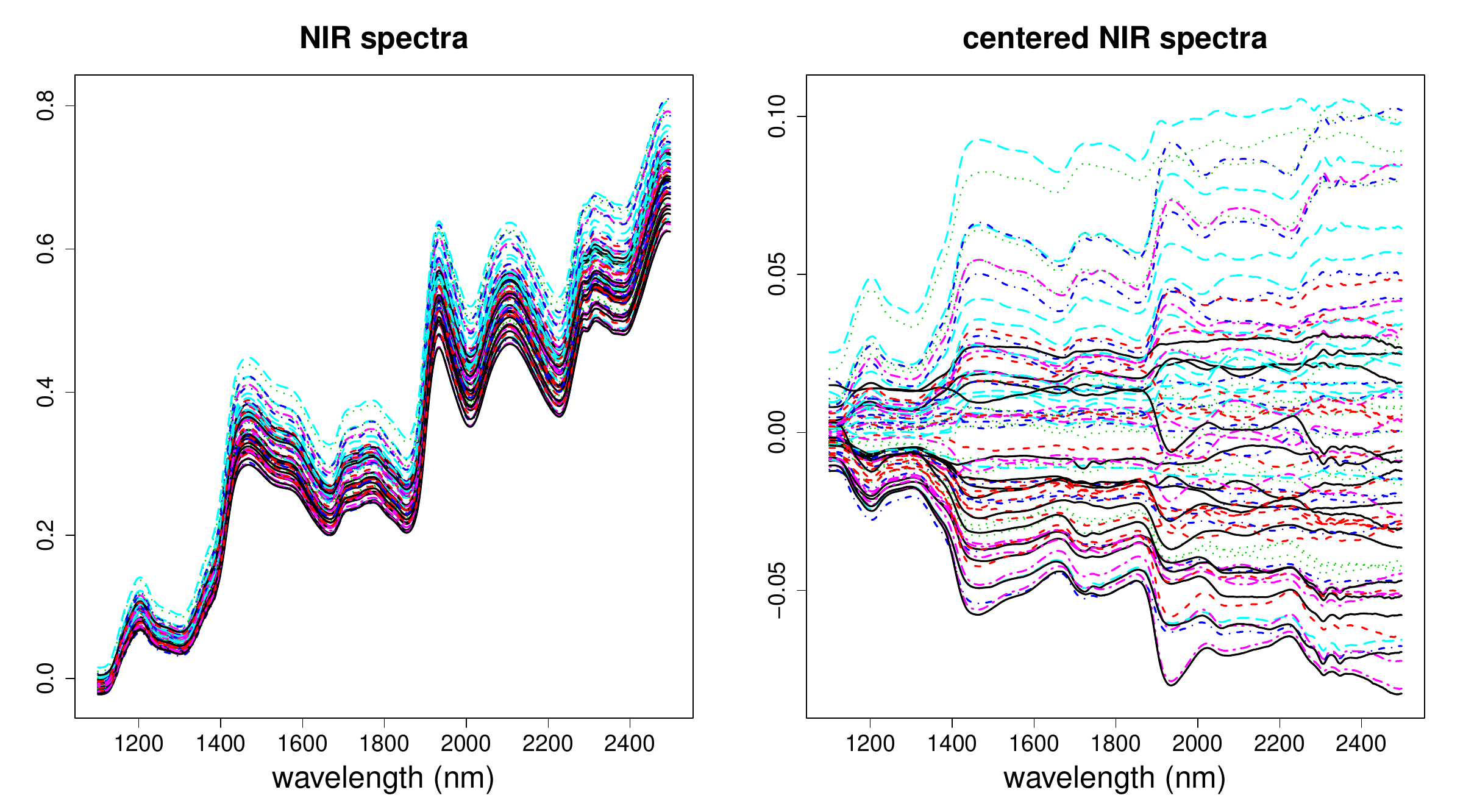}
 \end{center}
    \caption{\label{fig_81} \baselineskip=10pt The original and centered NIR spectra curves of 80 samples in the corn data.}
\end{figure}

 People are interested in predicting the values of the four ingredients of corn based on the NIR spectrum curve. So we consider the functional linear regression model with the NIR spectrum curve as the single functional predictor and the four ingredients as the multivariate response. We plot all 80 predictor curves, together with their centered versions, in Figure \ref{fig_81}.  The four response variables are all centered and scaled to have variance one, and their histograms are shown in Figure S.2 in supplementary material.

We first apply all the methods in Section \ref{sec5.1} for Simulations 1 and 2 to this data set and compare their prediction performance. We perform 100 iterations, and in each iteration, we randomly split the total 80 observations into a training set with 60 observations and a test set with 20 observations. 
In addition to MSPE,  we also calculate the $R^2$ to evaluate the goodness of fit of each method using the following formula in each repeat,
\begin{align}
   R^2=1-\frac{\sum_{\ell=1}^{60}\|\widehat{\mathbf{Y}}_{{\rm fitted},\ell}-\mathbf{Y}_{{\rm train},\ell}\|_2^2}{\sum_{\ell=1}^{60}\|\mathbf{Y}_{{\rm train},\ell}-\overline{\mathbf{Y}}_{{\rm train}}\|_2^2},\label{3693}
\end{align} 
where the four-dimensional vector $\mathbf{Y}_{{\rm train},\ell}$ is the $\ell$-th observation of the response in the training set, and $\widehat{\mathbf{Y}}_{{\rm fitted},\ell}$ is the corresponding fitted value, $1\le \ell\le 60$. The average (and standard deviation) of the MSPEs over 100 repeats is 0.18 (0.03) for our method, 0.32 (0.08) for \textit{pfr},  0.43 (0.07) for \textit{fbasis}, 0.34 (0.05) for \textit{fpcr}, 0.22 (0.03)  for \textit{fpls}, 0.87 (0.14)  for \textit{fnp}, 1.01 (0.14) for \textit{gp-msof}, 0.31 (0.06) for \textit{wLasso} and  0.91 (0.13) for \textit{wwLasso}. The average (and standard deviation) of the $R^2$'s over 100 repeats is 91\% (0.8\%) for our method, 78\% (0.7\%) for \textit{pfr},  66\% (1\%) for \textit{fbasis}, 76\% (2\%) for \textit{fpcr}, 87\% (1\%)  for \textit{fpls}, 40\% (5\%)  for \textit{fnp}, 10\% (6\%) for \textit{gp-msof},  78\% (3\%) for \textit{wLasso} and 15\% (3\%) for \textit{wwLasso}.

Given a new predictor curve, in addition to the predicted values, it is also of interest to provide prediction intervals for the response to evaluate the uncertainty of prediction. With the proposed two-step procedure, we cannot provide explicit formula for the estimates of coefficient functions and their asymptotic variance matrix. So we propose to use the bootstrap method to calculate the prediction intervals, which also accounts for the variations due to tuning parameter selection. As an example, we use all the observations except the first four as the training data, and the first four observations as the test data. Using this training data, we will calculate the bootstrap prediction interval and estimate the expectation of the mean squared prediction error for each of the first four predictor curves. Specifically, we repeat the following procedure 1000 times. In each iteration,  we draw a bootstrap sample with replacement from the training data and with same size as the training data. We fit the model using the bootstrap sample and our method, and then calculate the predicted response for each observation in the test data. From the 1000 repeats, we obtain 1000 predicted values for each coordinate of the  response. Using the 2.5\% and 97.5\% percentiles of these predicted values as the lower and upper limits, we get a 95\% bootstrap prediction interval for the corresponding coordinate of the response. We show the predictor curves in the first four observations and provide the bootstrap prediction intervals in Figure S.3 and Table S.3  of supplementary material, respectively. The  expected mean squared prediction error for the first four observations are estimated as 0.13, 0.24, 0.04, 0.09, respectively.

\begin{figure}[h]
\begin{center}
 \includegraphics[height=3in,width=6.5in]{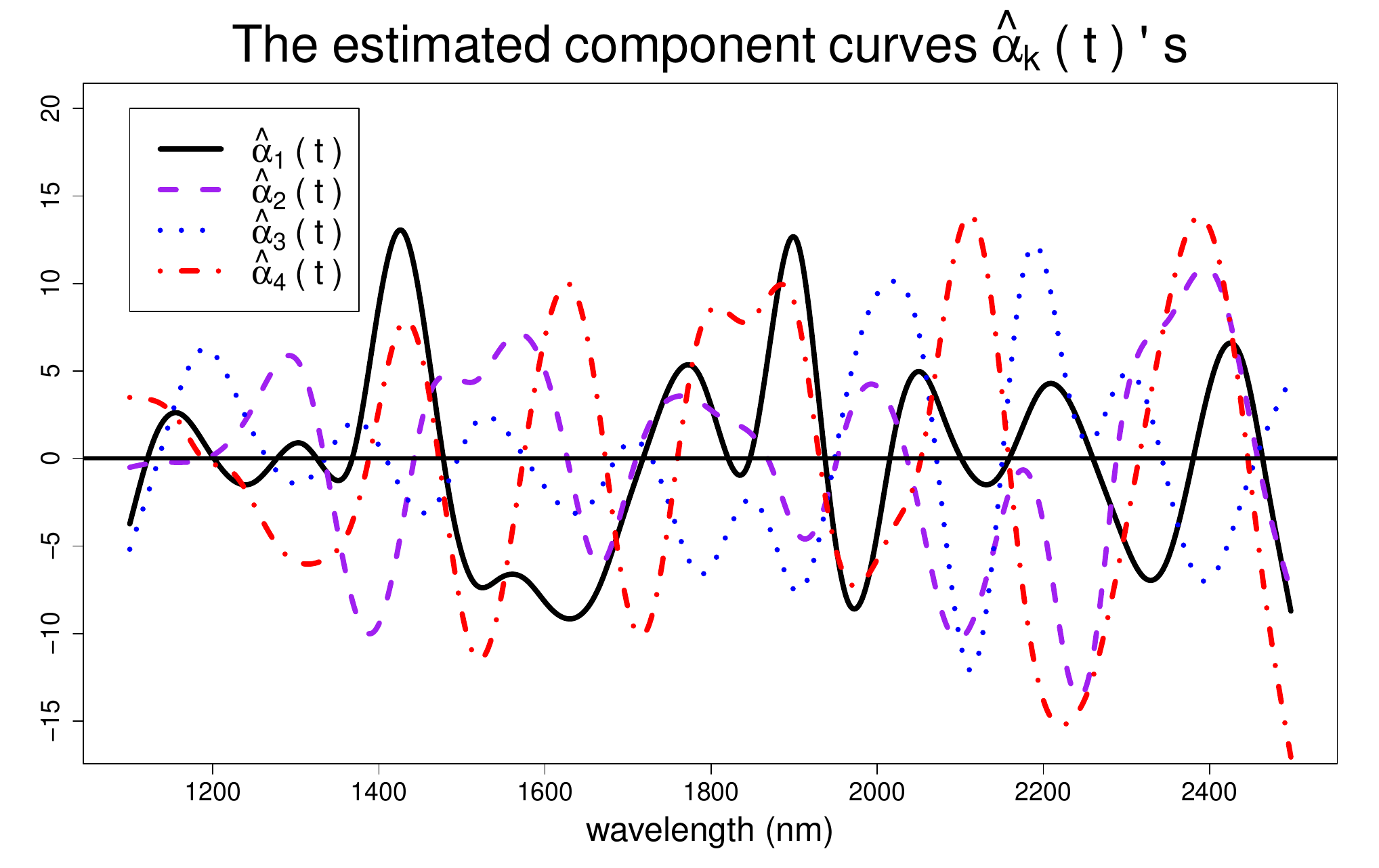}
\end{center}
    \caption{\label{fig_883} \baselineskip=10pt The estimates $\widehat{\alpha}_k(t)$, $1\le k\le 4$, for the corn data.}
\end{figure}

Finally, applying our method to all 80 observations, we obtain a model which is built upon four components. The ratio of the four estimators $\widehat{\sigma}^2_k$ of  $\sigma^2_k$, $1\le k\le 4$, is $0.51:0.34:0.14:0.01$. We plot the estimators $\widehat{\alpha}_k(t)$'s of the functions  $\alpha_k(t)$ for the four components in Figure \ref{fig_883}.

\subsection{Simultaneous EEG-fMRI data}\label{section_6.2}

We consider a simultaneous EEG-fMRI data set ({\it https://openfmri.org}, access number ds000116) from the project  ``{\it Auditory and Visual Oddball EEG-fMRI}''. This data set contains 102 observations recorded in 102 runs, each of which lasts 340 seconds. During each run, 125 total visual (or auditory) stimuli were presented and each of them lasts 200 ms followed by a $2\sim 3$ seconds intertrial interval. For the background and details of the experiments, we refer the reader to \citet{walz2013simultaneous}. In each observation, both fMRI time series and EEG time series are simultaneously recorded for the subject. We will extract multivariate responses from EEG data and functional predictors from fMRI data.  

The EEG data were recorded with sampling rate 1,000 Hz and has been preprocessed to remove gradient-artifact and be re-referenced to 34-channel electrode space  \citep{walz2013simultaneous}. So in each observation, the EEG data contains $34$ times series each of which has $340,000$ observation time points. It is known that the fluctuations of EEG power spectrum in different frequency bands are related to different brain activities. For example, the gamma frequency band (between 25Hz and 100Hz) is thought to be related to visual \citep{gray1989oscillatory, engel1991interhemispheric, fries2001modulation, womelsdorf2006gamma}, auditory \citep{brosch2002stimulus, edwards2005high},  and other sensory processing. For each observation, we calculate a gamma power spectrum using  ``{\sf Chronux}'' \citep{Chronux}, a popular software package implemented as a Matlab library for the analysis of neural data. The 34 EEG time series for 34 channels are combined to calculate a gamma power spectrum for each observation. Examples of gamma power spectrum for two observations are plotted in the top of Figure S.4 of supplementary material. There are many peaks in these spectrum curves and the numbers and positions of these peaks may vary for different observations. We identify ten common peaks and plot them in the bottom of Figure S.4 in supplementary material. We use the integral of each peak in the interval shown in Figure S.4 with length 0.2 HZ to measure its magnitude. The ten  magnitude values for the ten peaks form our ten-dimensional multivariate response. As the values of the magnitudes are highly skewed distributed, we  transform the responses using the function $\sqrt{\ln(1+u)}$, and scale the transformed values such that each coordinate of the ten-dimensional response vector has sample variance one. We draw the histograms for ten coordinates of the transformed response in Figure S.5 of supplementary material.  

The fMRI data contain 170 three-dimensional images obtained every 2 seconds. Each fMRI image is divided into $64\times 64\times 32=131,072$ voxels. Therefore, in each of the 102 observations, the fMRI data can be viewed as a collection of $131,072$ time series curves. We process these curves in two different ways and consider two different functional regression models.

\noindent \underline{\it (1). Using regions of interest (ROI) curves.} \\
We partition the brain voxels into 68 ROIs using the WFU PickAtlas Tools \citep{lancaster1997talairach, lancaster2000automated, maldjian2003automated} in Matlab SPM12 \citep{ashburner2014spm12} based on the TD Brodmann areas of human brain \citep{brodmann1909vergleichende}. We average the time series curves for all voxels in each ROI and use the $p=68$ averaged curves as functional predictors.

With 68 close to the number of functional predictors in Simulation 3, we apply all the methods used in Simulation 3 to this model. We perform 100 iterations. In each iteration, we randomly split the 102 observations into the training set with 80 observations and the test set with 22 observations. We calculate both the MSPE and the $R^2$, and summarize them in Table~\ref{eeg_1.tab}. The average MSPEs of our method with and without sparsity penalty are very close, and much lower than other methods. 
Among the 68 ROIs, 33 of them are selected in all 100 iterations by our method (shown in Figure S.6 and listed in Table S.4 of supplementary material).  The activities in those regions may be related to the fluctuations of spectrum at the ten peaks. For example, the Brodmann area (BA) 7 is the visuo-motor coordination and is involved in locating objects in space; the BA 8 includes the frontal eye fields and is involved in planning complex movements; the BAs 17 and 18 are respectively the primary and secondary visual cortex that processes visual information; the optic tract is a part of the visual system; and the Lateral Dorsal Nucleus is a relay center in the thalamus for the visual pathway.

\begin{table}[h]
\caption{\baselineskip=10pt The averages (and standard deviations) of MSPEs and $R^2$ in the first model for the simultaneous EEG-fMRI data.}
\vspace{5mm}
\label{eeg_1.tab}\centering
 \small\addtolength{\tabcolsep}{-3pt}
 {
\begin{tabular}{ |c|c|c|c|c|c|c|}
\hline
  &\textit{new.SpSm} & \textit{new.Sm}   & \textit{wLasso}   &\textit{wwLasso} & \textit{FPCA+gSCAD} & \textit{FPCA+gLasso} \\\hline\hline
MSPE & 0.29(0.09) &0.29 (0.07)& 1.02(0.16) & 1.02(0.16) & 1.12(1.08)  & 1.19(1.3)  \\\hline
 $R^2$ & 0.86(0.03) & 0.88(0.01) & 0.02(0.03) & 0.02(0.02) & 0.03(0.03)  & 0.04(0.04)   \\\hline
\end{tabular}}
\end{table}

\noindent \underline{\it (2). Using voxel curves.} \\
In the second model, we work on the voxel time serie curves directly. With a pre-screen procedure described in Section S.3.3 in supplementary material, we select $p=3559$ voxels and use their time series curves as our predictive curves. We show the positions of the 3559 voxels in the brain images in Figure S.7 and plot 200 among the 3559 predictive curves of one sample in Figure S.8 in Section S.3.3 of  supplementary material.  
 
Using the fMRI time series curves of the $3559$ voxels as functional predictors, we apply the methods in Simulation 4 to fit the model. Based on 100 iterations as in the study using ROI curves, the average (standard deviation) of the MSPEs are  0.176(0.029) for our method \textit{new.SmSp}, 1.023(0.175) for \textit{wLasso} and 1.044(0.186)  for \textit{wwLasso}.  Compared to the model using 68 ROIs, when using $p=3559$ voxels, the prediction error of our method decreases greatly. This implies that information is lost when taking average of the voxel time series in each ROI. Among the 3559 voxels, 49 of them are selected more than 90\% times, and 97 are selected more than 80\% times. These selected voxels are sparse and does not show a structured clustering. A possible way to improve the voxel selection
is to incorporate a group penalty in the simultaneous sparse-smooth penalty $\eqref{1003}$, where the group refers to the ROI information. This will be explored in future study.

\bibliographystyle{rss}

\end{document}